\documentclass[10pt,journal,a4paper]{IEEEtran}

\usepackage{color}
\usepackage{siunitx}
\DeclareSIUnit \dBm {dBm}
\DeclareSIUnit \dB {dB}
\DeclareSIUnit \Mbps {Mbps}
\DeclareSIUnit \Gbps {Gbps}
\DeclareSIUnit \mph {mph}
\usepackage{cite}      
\usepackage[caption=false, font=footnotesize]{subfig}
\usepackage[dvips]{graphicx}
\usepackage[latin1]{inputenc}
\usepackage{url}       
\usepackage{amsfonts}
\usepackage{amsmath}
\usepackage{times}
\usepackage{algpseudocode}
\usepackage{algorithm}
\usepackage{enumerate}
\usepackage{siunitx}
\usepackage{multirow}
\usepackage{tikz}
\usepackage{slashbox}

\newtheorem{assumption}{Assumption}[section]

\newtheorem{lemma}{Lemma}[section]
\newtheorem{theorem}{Theorem}[section]
\newtheorem{corollary}{Corollary}[section]
\newtheorem{remark}{Remark}[section]
\newtheorem{example}{Example}[section]

\newcolumntype{P}[1]{>{\centering\arraybackslash}p{#1}}
\newcolumntype{M}[1]{>{\centering\arraybackslash}m{#1}}

\begin{document}
\sisetup{range-phrase=--}
\sisetup{range-units=single}
\title{Modeling and Design of Millimeter-Wave Networks for Highway Vehicular Communication}

\author{Andrea Tassi, Malcolm Egan, Robert J. Piechocki and Andrew Nix
\thanks{
This work is partially supported by the \mbox{VENTURER} Project and \mbox{FLOURISH} Project, which are supported by Innovate UK under Grant Numbers 102202 and 102582, respectively.

A. Tassi, R. J. Piechocki and A. Nix are with the Department of Electrical and Electronic Engineering, University of Bristol, UK (e-mail: \{\mbox{a.tassi}, \mbox{r.j.piechocki}, \mbox{andy.nix}\}@bristol.ac.uk).

M. Egan is with the CITI Laboratory of the Institut National de Recherche en Informatique et en Automatique (INRIA), Universit{\'e} de Lyon, and Institut National de Sciences Apliqu{\'e}es (INSA) de Lyon, FR (e-mail: \mbox{malcom.egan}@inria.fr).}}

\maketitle

\begin{abstract}
Connected and autonomous vehicles will play a pivotal role in future Intelligent Transportation Systems (ITSs) and smart cities, in general. High-speed and low-latency wireless communication links will allow municipalities to warn vehicles against safety hazards, as well as support cloud-driving solutions to drastically reduce traffic jams and air pollution. To achieve these goals, vehicles need to be equipped with a wide range of sensors generating and exchanging high rate data streams. Recently, millimeter wave (mmWave) techniques have been introduced as a means of fulfilling such high data rate requirements. In this paper, we model a highway communication network and characterize its fundamental link budget metrics. In particular, we specifically consider a network where vehicles are served by mmWave Base Stations (BSs) deployed alongside the road. To evaluate our highway network, we develop a new theoretical model that accounts for a typical scenario where heavy vehicles (such as buses and lorries) in slow lanes obstruct Line-of-Sight (LOS) paths of vehicles in fast lanes and, hence, act as blockages. Using tools from stochastic geometry, we derive approximations for the Signal-to-Interference-plus-Noise Ratio (SINR) outage probability, as well as the probability that a user achieves a target communication rate (rate coverage probability). Our analysis provides new design insights for mmWave highway communication networks.
{In considered highway scenarios, we show that reducing the horizontal beamwidth from $90^\circ$ to $30^\circ$ determines a minimal reduction in the SINR outage probability (namely, $4 \cdot 10^{-2}$ at maximum). Also, unlike bi-dimensional mmWave cellular networks, for small BS densities (namely, one BS every $\SI{500}{\meter}$) it is still possible to achieve an SINR outage probability smaller than $0.2$.}
\end{abstract}


\section{Introduction}\label{sec:intro}
By 2020, fifty billion devices will have connectivity capabilities~\cite{cisco}. Among these, ten million vehicles equipped with on-board communication systems and with a variety of autonomous capabilities will be progressively rolled out. According to the National Highway Traffic Safety Administration (U.S. Department of Transportation) and the European Commission's Connected-Intelligent Transportation System (C-ITS) initiative~\cite{NHTSA,C_ITS}, connectivity will allow vehicles to engage with future ITS services, such as See-Through, Automated Overtake, High-Density Platooning, etc~\cite{5G-PPP}.

As identified by the European Commission's C-ITS initiative, the number of sensors mounted on each vehicle has increased. A typical sensor setup is expected to range from ultra-sound proximity sensors to more sophisticated camcorders and `Light Detection And Ranging' (LiDAR) systems~\cite{5G-PPP}. Currently, the number of on-board sensors are around $100$ units and this number is expected to double by 2020~\cite{RR0.3}. Ideally, the higher the number of on-board sensors, the ``smarter'' the vehicle. However, this holds true only if vehicles are able to exchange the locally sensed data~\cite{R2}. For instance, multiple LiDAR-equipped vehicles may approach a road hazard and share their real-time LiDAR data with incoming vehicles by means of the road-side infrastructure. This allows the approaching vehicles to compensate for their lack of sensor data (blind-spot removal) and, for instance, help smart cruise-control systems make decisions.
As such, there are strong constraints on LiDAR data delivery, which can be generated at rates up to \SI{100}{\Mbps}.
More generally, semi-autonomous and fully autonomous vehicles will require high rate and low latency communication links to support the applications envisaged by the 5G Infrastructure Public Private Partnership's (5G-PPP). These applications include the See-Through use case (maximum latency equal to \SI{50}{\milli\second}), which enables vehicles to share live video feeds of their onboard cameras to following vehicles. Other applications such as Automated Overtake and High-Density Platooning are also expected to require communication latencies smaller than \SI{10}{\milli\second}~\cite[Table~1]{5G-PPP}.

Recently, communication systems operating in the millimeter-wave (mmWave) range of the wireless spectrum have been proposed as a means of overcoming the rate and latency limitations of existing technologies~\cite{RR0,RR1}. In fact, currently commercialized mmWave systems can already ensure up to \SI{7}{\Gbps} and latencies smaller than \SI{10}{\milli\second}~\cite{C0}. Table~\ref{tab:technology_comparison} summarizes the general performance metrics of mmWave systems and compares them with the main technologies adopted to enable infrastructure-to-vehicle communications. Traditionally, ITSs rely on Dedicated Short-Range Communication (DSRC) standards, such as \mbox{IEEE 802.11p/DSRC} and ITS-G5/DSRC~\cite{R0,R3,R4,5771027}. Even though these technologies operate in a licensed band and ensure low communication latencies, their maximum realistic data rate hardly exceeds \SI{6}{\Mbps}~\cite{R0}. As such, several papers~\cite{7390802,6515060} suggest the adoption of 3GPP's Long Term Evolution-Advanced (LTE-A)~\cite{7849790,7335581}, which can guarantee higher communication rates. Nevertheless, the maximum supported data rate is limited to \SI{100}{\Mbps} and end-to-end latencies cannot go below \SI{100}{\milli\second}~\cite{R2}. {As a result, both DSRC and LTE-A cannot always meet the communication constraints dictated by \emph{delay and bandwidth sensitive} services that will be offered by future ITSs~\cite[Table~1]{5G-PPP}}.

In mmWave systems, both Base Stations (BSs) and users are equipped with large antenna arrays to achieve high array gains via beamforming techniques~\cite{rappaport2014millimeter}. As mmWave systems operate in the portion of the spectrum between \SI{30}{\giga\hertz} and \SI{300}{\giga\hertz}~\cite{C0}, mmWave links are highly sensitive to blockages. In particular, line-of-sight (LOS) communications are characterized by path loss exponents that tend to be smaller than $2.8$, while non-line-of-sight (NLOS) path loss exponents are at least equal to $3.8$~\cite{6387266}.
Due to of the the difficulty of beam alignment, commercial mmWave solutions cannot support user speeds greater than \SI{100}{\kilo\meter\per\hour}.
Despite this, research to cope with mobile users is gaining momentum~\cite{7156092}. For instance, in the UK, the main railway stakeholders are already trialing mmWave systems with enhanced beam searching techniques to provide broadband wireless connectivity onboard moving trains~\cite{mantra}. In addition, multiple research initiatives already regard mmWave systems as suitable to deploy 5G cellular networks~\cite{7474037,6932503,6515173}.

\begin{table}[t]
\renewcommand{\arraystretch}{1.3}
\centering
    \caption{Radio access solutions for vehicular communications.}
{\scriptsize\begin{tabular}{|M{1.2cm}|M{2.3cm}|M{1.8cm}|M{1.8cm}|}
\cline{2-4}
    \multicolumn{1}{c|}{} & \textbf{IEEE 802.11p/DSRC, ITS-G5/DSRC}~\cite{5771027} & {\textbf{LTE-A}~\cite{6515060}} & {\textbf{mmWave Systems}~\cite{C0}} \\ \hline
	\textbf{Frequency Band} & \SI{5.85}{\giga\hertz} - \SI{5.925}{\giga\hertz} & Spanning multiple bands in \SI{450}{\mega\hertz} - \SI{4.99}{\giga\hertz} & $\SI{28}{\giga\hertz}$, $\SI{38}{\giga\hertz}$, $\SI{60}{\giga\hertz}$ bands and E-band   \\ \hline
	\textbf{Channel Bandwidth} & \SI{10}{\mega\hertz} &  Up to \SI{100}{\mega\hertz} & \SI{100}{\mega\hertz}-\SI{2.16}{\giga\hertz} \\ \hline 
	\textbf{Bit Rate} & \SI{3}{\Mbps}-\SI{27}{\Mbps} & Up to \SI{1}{\Gbps} & Up to \SI{7}{\Gbps} \\ \hline
	\textbf{Latency} & $\leq \SI{10}{\milli\second}$ & \SI{100}{\milli\second}-\SI{200}{\milli\second} & $\leq \SI{10}{\milli\second}$ \\ \hline 
	\textbf{Mobility Support} & $\leq$ \SI{130}{\kilo\meter\per\hour} & $\leq$ \SI{350}{\kilo\meter\per\hour} & $\leq$ \SI{100}{\kilo\meter\per\hour} \\ \hline
	\end{tabular}}
    \label{tab:technology_comparison}
\end{table}

In this paper, we consider a typical road-side infrastructure for ITSs~\cite{hasan2012intelligent}. {In particular, the infrastructure-to-vehicle communications required by ITS services are handled by a dedicated network of BSs placed on dedicated antenna masts and or other street furniture, typically on both sides of the road~\cite{citeulike:3331930,Trullols2010432}.} We deal with a highway system where vehicles receive high data rate streams transmitted by mmWave BSs, although we do not consider scenarios where there is no roadside deployment of BSs where vehicle-to-vehicle communication technologies may provide a more effective solution.
A key feature of our highway system is that vehicles with different sizes are likely to drive along the same set of lanes. In a left-hand traffic system, any slow vehicle (such as double decker buses or lorries) typically travels in the outermost lanes of the highway, while the other vehicles tend to drive along the innermost lanes. If a larger vehicle drives between a user and its serving BS, the BS is no longer in LOS. In other words, large vehicles may act as communication blockages.

We develop a new framework to analyze and design mmWave communication systems. The original contributions of this paper are summarized as follows:
\begin{itemize}
\item We propose the first theoretical model to characterize the link budget requirements of mmWave networks providing downlink connectivity to highway vehicles where communications are impaired by large vehicles acting as communication blockages. Specifically, we offer design insights that take into account how the BS and blockage densities impact the user achievable data rate.
\item We show that the performance of mmWave highway networks can be well approximated by our theoretical model, which assumes that both BS and blockage positions are governed by multiple time-independent mono-dimensional PPPs.
{Traditional vehicular models assume that BSs are equally spaced~\cite{RW0} -- thus making them incapable of describing irregular BS deployments. In addition, the impact of blockages is either not considered~\cite{RW2} or the blockage positions are deterministically known~\cite{RW1}, which makes the latter kind of models suitable to be included in large-scale network simulators but also makes them analytically intractable. On the other hand, the proposed model is analytically tractable and allows us to predict user performance in scenarios characterized by different BS densities, traffic intensities, antenna gain and directivity without assuming the BS and blockage positions being known in advance.}
\item {Our numerical validation demonstrates that the proposed theoretical model is accurate and provides the following design insights: (i) a smaller antenna beamwidth does not necessarily reduce the Signal-to-Interference-plus-Noise Ratio (SINR) outage probability, and (ii) a reduced SINR outage probability in highway mmWave networks can be achieved even by low-density BS deployments, for a fixed probability threshold.}
\end{itemize}

The remainder of the paper is organized as follows. Section~\ref{sec.RW} discusses the related work on mmWave and vehicular communication systems. Section~\ref{sec:SM} presents our mmWave communication system providing downlink coverage in highway mmWave networks. 
We evaluate the network performance in terms of the SINR outage and rate coverage probabilities, which are derived in Section~\ref{sec.sinr}. Section~\ref{sec.nr} validates our theoretical model. In Section~\ref{sec.cl}, we conclude and outline avenues of future research.

\section{Related Work}\label{sec.RW}
\begin{table*}[t]
\centering
\caption{Related Works on mmWave Systems and Vehicular Communications.}
\label{tab.refs}
{\scriptsize
\begin{tabular}{|M{0.5cm}|M{1.5cm}|M{3cm}|M{3cm}|M{3.5cm}|M{2cm}|M{1cm}|M{2cm}|}
\hline \textbf{Ref.}  & \textbf{Radio Access Technology} & \textbf{Network Topology} & \textbf{Channel and Path Loss Models} & \textbf{Mobility} & \textbf{Communication Blockages}\\
\hline \cite{RR0}      & mmWave & Vehicle-to-vehicle, Vehicle-to-infrastructure & Based on ray-tracing & Vehicles moving on urban roads & Not analytically investigated\\
\hline \cite{6932503}  & mmWave & Dense cellular network & Nakagami small-scale fading; BGG path loss model & Static blockages & Buildings \\
\hline \cite{C0}       & mmWave & Dense cellular system & Based on measurements & Static blockages & Buildings \\
\hline \cite{C1}       & mmWave & Network backhauling & Constant small-scale fading (i.e., the square norm of the small-scale fading contribution is equal to $1$) & Static blockages & None \\
\hline \cite{C2}       & mmWave & Cellular network with self-backhauling & Constant small-scale fading; BGG path loss model & Static blockages & Buildings \\
\hline \cite{7370940}  & mmWave & Co-operative cellular network & Nakagami for the signal, Rayleigh for the interference contribution; BGG path loss model & Static blockages & Buildings \\
\hline \cite{7010535}  & mmWave & Cellular network with self-backhauling & Based on measurements & Static blockages & Indoor objects\\
\hline \cite{7105406}  & mmWave & Multi-tier cellular network & Constant small-scale fading; Probabilistic path loss model & Static blockages & Buildings \\
\hline \cite{RW0}      & DSRC   & Vehicle-to-vehicle, Vehicle-to-infrastructure & Coverage-based (i.e., no packet errors from nodes within the radio range) & Vehicles moving on urban roads & None \\
\hline \cite{RW2}      & DSRC   & Vehicle-to-vehicle, Vehicle-to-infrastructure & Rice small-scale fading & Vehicles moving on urban roads & None \\
\hline \cite{RW1}      & DSRC   & Vehicle-to-vehicle & Obstacle-based channel and path loss models & Vehicles moving on a highway & Vehicles \\
\hline \cite{RW4}      & DSRC   & Vehicle-to-vehicle & Rayleigh small-scale fading & Vehicles moving on a highway & None \\
\hline
\end{tabular}
}
\end{table*}

As summarized in Table~\ref{tab.refs}, over the past few years, mmWave systems have been proposed as a viable alternative to traditional wireless local area networks~\cite{C0} or as a wireless backhauling technology for BSs of the same cellular network~\cite{C1,C2}. Furthermore, mmWave technology has also been considered for deploying dense cellular networks characterized by high data rates~\cite{6932503,7010535,7105406}. With regards to the vehicular communication domain, J.~Choi~\textit{et al.}~\cite{RR0} pioneered the application of the mmWave technology to partially or completely enable ITS communications. A mmWave approach to ITS communications is also being supported by the European Commission~\cite{C_ITS}.

As both the BS deployment and vehicle locations differ over both time and in different highway regions, any highway network model must account for these variations. In this setting, stochastic geometry provides a means of characterizing the performance of the system by modeling BS locations via a spatial process, such as the Poisson Point Process (PPP)~\cite{SGC0}.
Generally, PPP models for wireless networks are now a well-established methodology~\cite{1542405,Lee2013,SGC0,6524460,6497002,RR1}; however, there are challenges in translating standard results into the context of mmWave networks for road-side deployments due to the presence of NLOS links resulting from blockages~\cite{6932503}.
In particular, the presence of blockages has only been addressed in the context of mmWave cellular networks in urban and suburban environments that are substantially different to a highway deployment~\cite{6932503}. In particular, in mmWave cellular networks: (i) the positions of BSs follow a bi-dimensional PPP, and (ii) the positions of blockages are governed by a stationary and isotropic process. Even though this is a commonly accepted assumption for bi-dimensional cellular networks~\cite{SGC0}, this is not satisfied by highway scenarios, where both blockages and BS distributions are clearly not invariant to rotations or translations. With regards to Table~\ref{tab.refs}, the path loss contribution of blockages has either been modeled by means of the Boolean Germ Grain (BGG) principle (i.e., only the BSs within a target distance are in LOS) or in a probabilistic fashion (i.e., a BS is in LOS/NLOS with a given probability). To the best of our knowledge, no models for road-side mmWave BS deployment accounting for vehicular blockages have been proposed to date.

Given the simplicity of their topology and their high level of automation, highway scenarios have been well investigated in the literature~\cite{RW0,RW2,RW1}. In particular,~\cite{RW0} addresses the issue of optimizing the density of fixed transmitting nodes placed at the side of the road, with the objective of maximizing the stability of reactive routing strategies for Vehicular Ad-Hoc Networks (VANETs) based on the IEEE 802.11p/DSRC stack.
Similar performance investigations are conducted in~\cite{RW2} where a performance framework jointly combining physical and Media Access Control (MAC) layer quality metrics is devised.
In contrast to~\cite{RW0} and~\cite{RW2},~\cite{RW1} addresses the issue of blockage-effects caused by large surrounding vehicles; once more,~\cite{RW1} strictly deals with IEEE 802.11p/DSRC communication systems. The proposal in~\cite{RW0,RW2,RW1} is not applicable for mmWave highway networks as the propagation conditions of a mmWave communication system are not comparable with those characterizing a system operating between \SI{5.855}{\giga\hertz} and \SI{5.925}{\giga\hertz}.
Another fundamental difference with between mmWave and IEEE 802.11p/DSRC networks is the lack of support for antenna arrays capable of beamforming as the IEEE 802.11p/DSRC stack is restricted to omnidirectional or non-steering sectorial antennas.

Highway networks have also been studied using stochastic geometry. In particular, M. J. Farooq~\textit{et al.}~\cite{RW4} propose a model for highway vehicular communications that relies on the physical and MAC layers of an IEEE 802.11p/DSRC or ITS-G5/DSRC system. In particular, the key differences between this paper and~\cite{RW4} are: (i) the focus on multi-hop LOS inter-vehicle communications and routing strategies while our paper deals with one-hop infrastructure-to-vehicle coverage issues, and (ii) the adoption of devices with no beamforming capabilities while beamforming is a key aspect of our mmWave system.

\section{System Model and Proposed BS-Standard User Association Scheme}\label{sec:SM}
\begin{figure}[t]
\vspace{-3mm}
\centering
\includegraphics[width=1\columnwidth]{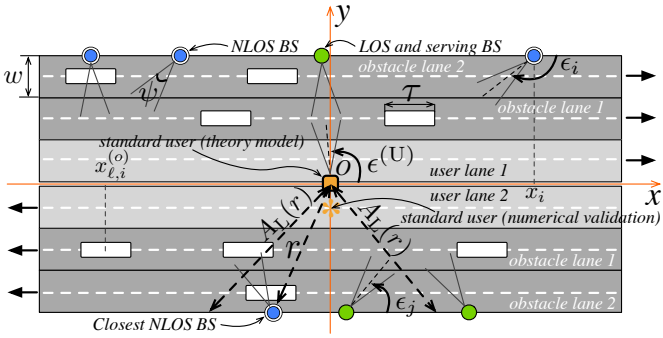}
\caption{Considered highway system model, composed of $N_o = 2$ obstacle lanes in each traffic direction.}\label{fig.SM}
\vspace{-4mm}
\end{figure}

We consider a system model where mmWave BSs provide network coverage over a section of a highway, illustrated in Fig.~\ref{fig.SM}. The goal of our performance model is to characterize the coverage probability of a user surrounded by several moving blockages (i.e., other vehicles) that may  prevent a target user to be in LOS with the serving BS. Without loss of generality, we consider the scenario where vehicles drive on the left-hand side of the road\footnote{The proposed theoretical framework also applies to road systems where drivers are required to drive on the right-hand side of the road.}. For clarity, Table~\ref{tab.not} summarizes the symbols commonly used in the paper. In order to gain insight into the behavior of the model, we make the following set of assumptions.

\begin{assumption}[Road Layout]\label{ass.roadL}
We assume that the whole road section is constrained within two infinitely long parallel lines, the \emph{upper} and \emph{bottom sides} of the considered road section. Vehicles flow along multiple parallel lanes in only two possible directions: West-to-East (for the upper-most lanes) and East-to-West (for the lowermost lanes). Each lane has the same width $w$. For each direction, there are $N_o$ obstacle lanes and one user lane closer to the innermost part of the road. The closer a lane is to the horizontal symmetry axis of the road section, the more the average speed is likely to increase -- thus, the large/tall vehicles are assumed to drive along obstacle lanes most of the time. Vehicles move along the horizontal symmetry axis of each lane. We use a coordinate system centered on a point on the line separating the directions of traffic. The upper side of the road intercepts the $y$-axis of our system of coordinates at the point $(0, w(N_o + 1))$, while the bottom side intercepts at $(0, -w(N_o + 1))$. 
\end{assumption}

In the following sections, we will focus on characterizing the performance of the downlink phase of a mmWave cellular network providing connectivity to the vehicles flowing in the high speed lanes of the considered model, which is challenging. In fact, communication links targeting users in the high speed lanes are impacted by the largest number of communication blockages (namely, large vehicles) flowing on on the outer road lanes. In addition, we adopt the standard assumption of the BSs being distributed according to a PPP.

\begin{table}[t!]
\centering
\caption{Commonly used notation.}
\label{tab.not}
{\scriptsize
\begin{tabular}{|c|p{6.5cm}|}
\hline $N_o$  & Number of obstacle lanes per driving direction\\
\hline $w$  & Width of a road lane\\
\hline $\lambda_{\mathrm{BS}}$  & Density of the PPP $\Phi_{\mathrm{BS}}$ of $x$-components of the BS locations on the road\\
\hline $\lambda_{O,\ell}$  & Density of the PPP $\Phi_{\mathrm{O},\ell}$ of $x$-components of the blockages on the $\ell$-th obstacle lane\\
\hline $\tau$  & Footprint segment of each blockage\\
\hline $p_\mathrm{L}$, $p_\mathrm{N}$  & Approximated probabilities of a BS being in LOS or NLOS with respect to the standard user, respectively\\
\hline $\lambda_\mathrm{L}$, $\lambda_\mathrm{N}$  & Densities of the PPPs of the $x$-components of LOS and NLOS BSs, respectively\\
\hline $\ell(r_i)$  & Path loss component associated with the $i$-th BS\\
\hline $A_\mathrm{L}$  & Assuming the standard user connects to a NLOS BS at a distance $r$, it follows that there are no LOS BSs at a distance less or equal to $A_{\mathrm{L}}(r)$\\
\hline $A_\mathrm{N}$  & Assuming the standard user connects to a LOS BS at a distance $r$, it follows that there are no NLOS BSs at a distance less or equal to $A_{\mathrm{N}}(r)$\\
\hline $\mathrm{P}_\mathrm{L}$, $\mathrm{P}_\mathrm{N}$  & Probabilities that the standard user connects to a LOS or a NLOS BS, respectively\\
\hline $G_{TX}$, $G_{RX}$  & Maximum transmit and receive antenna gains, respectively\\
\hline $g_{TX}$, $g_{RX}$  & Minimum transmit and receive antenna gains, respectively\\
\hline $\mathcal{L}_{\mathrm{I}_{\mathrm{S,E}},\mathbb{E}_1}(s)$  & Laplace transform of the interference component determined by BSs placed on the upper ($\mathrm{S} = \mathrm{U}$) or bottom side ($\mathrm{S} = \mathrm{B}$) of the road that are in LOS ($\mathrm{E} = \mathrm{L}$) or in NLOS ($\mathrm{E} = \mathrm{N}$) to the standard user, conditioned on the serving BS being in LOS ($\mathbb{E}_1 = \mathrm{L}$) or NLOS ($\mathbb{E}_1 = \mathrm{N}$)\\
\hline $\mathcal{L}_\mathrm{I,\mathbb{E}_1}(s)$  & Laplace transform of the interference $\mathrm{I}$, given that the standard user connects to a LOS BS ($\mathbb{E}_1 = \mathrm{L}$) or NLOS BS ($\mathbb{E}_1 = \mathrm{N}$)\\
\hline $\mathrm{P}_\mathrm{T}(\theta)$  & SINR outage probability as a function of SINR threshold $\theta$\\
\hline $\mathrm{R}_\mathrm{C}(\kappa)$  & Rate coverage probability as a function of target rate $\kappa$\\
\hline
\end{tabular}
}\vspace{-4mm}
\end{table}

\begin{assumption}[BS Distribution]\label{ass.BSdist}
Let $\Phi_{\mathrm{BS}} = \{x_i\}_{i = 1}^b$ be the one-dimensional PPP, with density $\lambda_{\mathrm{BS}}$ of the \mbox{$x$-components} of the BS locations on the road. We assume that BSs are located along with the upper and bottom sides of the road section. In particular, the $i$-th BS lies on the upper or bottom sides with a probability equal to $q = 0.5$. In other words, the $y$-axis coordinate of the $i$-th BS is defined as $y_i = w(-2\mathbb{B}_q+1)(N_o + 1)$, where $\mathbb{B}_q$ is a Bernoulli random variable with parameter $q$.
\end{assumption}

By Assumption~\ref{ass.BSdist} and from the independent thinning theorem of PPP~\cite[Theorem 2.36]{haenggi2013stochastic}, it follows that the \mbox{$x$-axis} coordinates of the BSs at the upper and bottom sides of the road form two independent PPPs with density $0.5 \cdot \lambda_{\mathrm{BS}}$.

\begin{assumption}[Blockage Distribution]\label{ass.BlDist}
We assume that the \mbox{$\ell$-th} obstacle lane  on a traffic direction and the coordinates $(x^{\mathrm{(o)}}_{\ell,i}, y^{\mathrm{(o)}}_{\ell,i})$ of blockage $i$, $x^{\mathrm{(o)}}_{\ell,i}$ belongs to a one-dimensional PPP $\Phi_{\mathrm{O},\ell}$ with density $\lambda_{\mathrm{o},\ell}$, for $\ell \in \{1, \ldots, N_o\}$~\cite{RW4}. The term $y^{\mathrm{(o)}}_{\ell,i}$ is equal to $w\ell$, or $-w\ell$, depending on whether we refer to the West-to-East or East-to-West direction, respectively. We assume that the density of the blockages of lane $\ell$ in each traffic direction is the same. Each blockage point is associated with a segment of length $\tau$, centered on the position of the blockage itself and placed onto the horizontal symmetry axis of the lane (hereafter referred to as the ``footprint segment''). Obstacles can be partially overlapped and the blockage widths and heights are not part of our modeling. The presence of large vehicles in the user lanes is regarded as sporadic hence, it is ignored.\label{ass.blockageDistribution}
\end{assumption}

From Assumption~\ref{ass.BlDist}, given a driving direction, we observe that the blockage density of each obstacle lane can be different. This means that our model has the flexibility to cope with different traffic levels per obstacle lane; namely, the larger the traffic density, the larger the traffic intensity. {In a real (or simulated) scenario, the obstacle density $\lambda_{\mathrm{o},\ell}$ of a road section is function of the mobility model, which in turn depends on the vehicle speed, maximum acceleration/deceleration, etc. Form a logic point of view, in our theoretical model, we observe that at the beginning of a time step, process $\Phi_{\mathrm{O},\ell}$ is sampled and a new blockage position is extracted, for $\ell = 1, \ldots, N_o$. In Section~\ref{sec.nr}, we will show that the considered PPP-based mobility model provides a tight approximation of the investigated network performance, in the case of blockages moving according to a Krauss car-following mobility model~\cite{Kanagaraj2013390}.}

Our primary goal consists of characterizing the SINR outage and rate coverage probability of users located on the user lanes, as these are the most challenging to serve due to the fact that vehicles in the other lanes can behave as blockages. For the sake of tractability, our theoretical model tractable will consider the service of a \emph{standard user} placed at the origin $O=(0,0)$ of the axis.

\vspace{-2mm}\subsection{BS-Standard User Association and Antenna Model}\label{sec.LN}
Since vehicles in the slow lanes can block a direct link between the standard user and each BS, it is necessary to distinguish between BSs that are in LOS with the standard user and those that are in NLOS.
{BS $i$ is said to be in LOS if the footprint of any blockage does not intersect with the ideal segment connecting the standard user and BS $i$. The probability that BS $i$ is in LOS is denoted by $p_{i,\mathrm{L}}$.
We assume that the blockages are of length $\tau$, illustrated in Fig.~\ref{fig.SM}. In the case that the ideal segment connecting BS $i$ to the standard user intersects with one or more footprint segments, BS $i$ is in NLOS (this occurs with probability $p_{i,\mathrm{N}}$) and the relation $p_{i,\mathrm{N}} = 1 - p_{i,\mathrm{L}}$ holds. For generality, we also assume that signals from NLOS BSs are not necessarily completely attenuated by the blockages located in the far-field of the antenna systems. This can happen when the main lobe of the antenna is only partially blocked and because of signal diffraction~\cite{Ant1,Ant2}.} By Assumption~\ref{ass.blockageDistribution}, we observe that the probability $p_{i,\mathrm{E}}$ for $\mathrm{E} \in \{\mathrm{L}, \mathrm{N}\}$ of BS $i$ being in LOS ($\mathrm{E} = \mathrm{L}$) or NLOS ($\mathrm{E} = \mathrm{N}$) depends on the distance from $O$. This is due to the fact that the further the BS is from the user, the further away the center of an obstacle footprint segment needs to be to avoid a blockage.

Consider Assumption~\ref{ass.blockageDistribution} and the set of points where the segment connecting $O$ with BS $i$ intersects the symmetry axis of each obstacle lane. We approximate $p_{i,\mathrm{L}}$ with the probability $p_{\mathrm{L}}$ that no blockages are present within a distance of $\tau/2$ on either side of the ray connecting the user to BS $i$. Hence, our approximation is independent of the distance from BS $i$ to $O$ and $p_{i,\mathrm{L}}$ can be approximated independently of $i$ as follows: 
\begin{equation}
p_{\mathrm{L}} \cong \prod_{\ell=1}^{N_o} e^{-\lambda_{\mathrm{o},\ell}\tau}, \label{eq.losNlosApp}
\end{equation}
while $p_{i,\mathrm{N}}$ is approximated as $p_{\mathrm{N}} = 1 - p_{\mathrm{L}}$. Observe also that the term $e^{-\lambda_{\mathrm{O},\ell}\tau}$ is the void probability of a one-dimensional PPP of density $\lambda_{\mathrm{o},\ell}$~\cite{haenggi2013stochastic}.

Using the approximation in~\eqref{eq.losNlosApp} and invoking the independent thinning theorem of PPP, it follows that the PPP of the LOS BSs $\Phi_{\mathrm{L}} \subseteq \Phi_{\mathrm{BS}}$ and of the NLOS BS $\Phi_{\mathrm{N}} \subseteq \Phi_{\mathrm{BS}}$ are independent and with density $\lambda_{\mathrm{L}} = p_{\mathrm{L}} \lambda_{\mathrm{BS}}$ and $\lambda_{\mathrm{N}} = p_{\mathrm{N}} \lambda_{\mathrm{BS}}$, respectively. In addition, the relation $\Phi_{\mathrm{L}} \cap \Phi_{\mathrm{N}} = \emptyset$ holds.

Consider the $i$-th BS at a distance $r_i = \sqrt{x_i^2 + y_i^2}$ from the standard user. The indicator function $\mathbf{1}_{i,\mathrm{L}}$ is equal to one if BS $i$ is in LOS with respect to the standard user, and zero otherwise. The path loss component $\ell(r_i)$ impairing the signal transmitted by BS $i$ and received by the standard user is defined as follows:
\begin{equation}
\ell(r_i) = \mathbf{1}_{i,\mathrm{L}} C_\mathrm{L} r_i^{-\alpha_\mathrm{L}} + (1-\mathbf{1}_{i,\mathrm{L}}) C_\mathrm{N} r_i^{-\alpha_\mathrm{N}}\label{eq.PL}
\end{equation}
where $\alpha_\mathrm{L}$ and $\alpha_\mathrm{N}$ are the path loss exponents, while $C_\mathrm{L}$ and $C_\mathrm{N}$ are the path loss intercept factors in the LOS and NLOS cases, respectively.
{Terms $C_\mathrm{L}$ and $C_\mathrm{N}$ can either be the result of measurements of being analytically derived by a free space path loss model; the intercept factors are essential to capture the path loss component at a target transmitter-receiver distance, which for practical mmWave system is equal to $\SI{1}{\meter}$~\cite{7481506}.}
From Assumption~\ref{ass.roadL}, we remark that relation $r_i \geq w(N_o + 1)$ holds. Hence, for typical values of road lane widths, path loss intercept factors and exponents, relation $w(N_o + 1) \geq \max\{C_\mathrm{L}^{\frac{1}{\alpha_\mathrm{L}}},C_\mathrm{N}^{\frac{1}{\alpha_\mathrm{N}}}\}$ holds as well. This ensures that both $C_\mathrm{L} r_i^{-\alpha_\mathrm{L}}$ and $C_\mathrm{N} r_i^{-\alpha_\mathrm{N}}$ are less than or equal to $1$.

\begin{assumption}[BS-Standard User Association]\label{ass.BSTUass}
In our system model, the standard user has perfect channel state information and always connects to the BS with index $i^*$, which is characterized by the minimum path loss component, i.e., $i^* = \displaystyle\arg\max_{i = 1, \ldots, b}\{\ell(r_i)\}$.
\end{assumption}

We assume that the standard user connects to a NLOS BS at a distance $r$. Since $w(N_o + 1) \geq \max\{C_\mathrm{L}^{\frac{1}{\alpha_\mathrm{L}}},C_\mathrm{N}^{\frac{1}{\alpha_\mathrm{N}}}\}$, it follows that there are no LOS BSs at a distance less than or equal to $A_{\mathrm{L}}(r)$, defined as:
\begin{equation}
A_{\mathrm{L}}(r) = \max\left\{w(N_o + 1),\left[\frac{C_{\mathrm{N}}}{C_{\mathrm{L}}} r^{-\alpha_{\mathrm{N}}}\right]^{-\frac{1}{\alpha_{\mathrm{L}}}}\right\}. \label{eq.Al}
\end{equation}
We observe that $A_{\mathrm{L}}(r)$ is the distance from $O$ for which the path loss component associated with a LOS BS is equal to the path loss component associated with a NLOS BS at a distance $r$. In a similar way, we observe that if the standard user connects to a LOS BS at a distance $r$ from $O$, it follows that there are no NLOS BSs at a distance smaller than or equal to $A_{\mathrm{N}}(r)$, defined as:
\begin{equation}
A_{\mathrm{N}}(r) = \max\left\{w(N_o + 1),\left[\frac{C_{\mathrm{L}}}{C_{\mathrm{N}}} r^{-\alpha_{\mathrm{L}}}\right]^{-\frac{1}{\alpha_{\mathrm{N}}}}\right\}. \label{eq.An}
\end{equation}
We observe that definitions~\eqref{eq.Al} and~\eqref{eq.An} prevent $A_{\mathrm{L}}(r)$ and $A_{\mathrm{N}}(r)$ to be smaller than the distance $w(N_o+1)$ between $O$ and a side of the road.

The standard user will always connect to one BS at a time, which is either the closest LOS or the closest NLOS BS. This choice is made by the standard user according to Assumption~\ref{ass.BSTUass}. In particular, is closest LOS BS is at a distance greater than $\mathrm{A_{L}}(r)$, the BS associated with the smallest path loss component is the closest NLOS BS, which is at a distance $r$ to the standard user. Those facts allow us to prove the following lemma.

\begin{lemma}\label{lem.pdf}
Let $\mathrm{d}_{\mathrm{L}}$ and $\mathrm{d}_{\mathrm{N}}$ be the random variables expressing the distance to the closest LOS and NLOS BSs from the perspective of the standard user, respectively. The Probability Density Function (PDF) of $\mathrm{d}_{\mathrm{L}}$ can be expressed as:
\begin{equation}
f_\mathrm{L}(r) = \frac{2 \lambda_{\mathrm{L}} r}{b(r)} e^{-2 \lambda_{\mathrm{L}} b(r)}, \label{eq.pdfL}
\end{equation}
while the PDF of $\mathrm{d}_{\mathrm{N}}$ can be expressed as
\begin{equation}
f_\mathrm{N}(r) = \frac{2 \lambda_{\mathrm{N}} r}{b(r)} e^{-2 \lambda_{\mathrm{N}} b(r)}, \label{eq.pdfN}
\end{equation}
where $b(r) = \sqrt{r^2 - w^2(N_o+1)^2}$.
\end{lemma}
\begin{IEEEproof}
Considering the LOS case, the proof directly follows from the expression of the PDF of the distance of the closest point to the origin of the axis in a one-dimensional PPP with density $\lambda_{\mathrm{L}}$, which is $f_\mathrm{L}(t) = 2 \lambda_{\mathrm{L}} e^{-2 \lambda_{\mathrm{L}} t}$~\cite[Eq.~(2.12)]{haenggi2013stochastic}. By applying the change of variable $t \leftarrow b(r)$ we obtain~\eqref{eq.pdfL}. With similar reasoning, it is also possible to prove~\eqref{eq.pdfN}.
\end{IEEEproof}

Using Lemma~\ref{lem.pdf},~\eqref{eq.Al} and~\eqref{eq.An}, the following lemma holds.
\begin{lemma}\label{lem.serv}
The standard user connects to a NLOS BS with probability
\begin{equation}
\mathrm{P}_{\mathrm{N}} = \int_{w(N_o+1)}^{\infty} f_\mathrm{N}(r) e^{-2 \lambda_{\mathrm{L}} b\left(A_{\mathrm{L}}(r)\right)} \,dr. \label{eq.Pn}
\end{equation}
On the other hand, the standard user connects to a LOS BS with probability
\begin{equation}
\mathrm{P}_{\mathrm{L}} = \int_{w(N_o+1)}^{\infty} f_\mathrm{L}(r) e^{-2 \lambda_{\mathrm{N}} b\left(A_{\mathrm{N}}(r)\right)} \,dr = 1 - \mathrm{P}_{\mathrm{N}}. \label{eq.Pl}
\end{equation}
\end{lemma}
\begin{IEEEproof}
Consider the event that the standard user connects to a NLOS BS, which is at a distance $r$ from $O$. This event is equivalent to have all the LOS BSs at a distance greater than or equal to $A_{\mathrm{L}}(r)$. From~\eqref{eq.pdfL}, it follows that $\mathbb{P}[\mathrm{d}_{\mathrm{L}} \geq A_{\mathrm{L}}(r) \,\,\text{and}\,\, \mathrm{d}_{\mathrm{N}} = r]$ is equal to $e^{-2 \lambda_{\mathrm{L}} b\left(A_{\mathrm{L}}(r)\right)}$. Then if we marginalize $\mathbb{P}[\mathrm{d}_{\mathrm{L}} \geq A_{\mathrm{L}}(r) \,\,\text{and}\,\, \mathrm{d}_{\mathrm{N}} = r]$ with respect to $r$, we obtain~\eqref{eq.Pn}. The same reasoning applies to the proof of~\eqref{eq.Pl}.
\end{IEEEproof}

The gain of the signal received by the standard user depends on the antenna pattern and beam steering performed by the BS and the user. Each BS and the standard user are equipped with antenna arrays capable of performing directional beamforming. To capture this feature, we follow~\cite{6932503} and use a sectored approximation to the array pattern. We detail the sectored approximation for our highway model in the following assumption.

\begin{assumption}[Antenna Pattern]\label{ass.Pattern}
The antenna pattern consists of a main lobe with beamwidth $\psi$ and a side lobe that covers the remainder of the antenna pattern. We assume that the gain of the main lobe is $G_{TX}$ and the gain of the sidelobe is $g_{TX}$. Similarly, the antenna pattern of the standard user also consists of main lobe with beamwidth $\psi$ and gain $G_{RX}$, and a side lobe with gain $g_{RX}$.
\end{assumption}

The antenna of each BS and the user can be steered as follows.

\begin{assumption}[BS Beam Steering]\label{ass.BSst}
Let $\epsilon_i$ be the angle between the upper (bottom) side of the road and the antenna boresight of BS $i$ (see Fig.~\ref{fig.SM}). We assume that $\epsilon_i$ takes values in \mbox{$\mathcal{G} = [\frac{\psi}{2}, 2\pi-\frac{\psi}{2}]$}. As such, the main lobe of each BS is always entirely directed towards the road portion constrained by the upper and bottom side. If the standard user connects to BS $i$, the BS steers its antenna beam towards the standard user. On the other hand, if the standard user is not connected to BS $i$, we assume that $\epsilon_i$ takes a value that is uniformly distributed in $\mathcal{G}$.
\end{assumption}

\begin{assumption}[Standard User Beam Steering]\label{ass.antennaPattern}
The angle $\epsilon^{(U)}$ between the positive $x$-axis and the boresight of the user beam is selected to maximize the gain of the received signal from the serving BS. We assume that $\epsilon^{(U)} \in [\frac{\psi}{2},\pi - \frac{\psi}{2}]$ or $\epsilon^{(U)} \in [\pi + \frac{\psi}{2},2\pi - \frac{\psi}{2}]$ if the user is served by a BS on the upper side or the bottom side of the road respectively. This assumption ensures that interfering BSs on the opposite side of the road are always received by a sidelobe, with gain $g_{RX}$. We also assume that the standard user directs its antenna beam towards the serving BS, which is then received with gain $G_{\mathrm{RX}}$.
\end{assumption}

\section{SINR Outage and Rate Coverage Characterization}\label{sec.sinr}
For the sake of simplifying the notation and without loss of generality, we assume that the BS with index $1$ is the BS that the standard user is connected to, while BSs $2, \ldots, b$ define the set of the interfering BSs. We define the SINR at the location of the standard user as follows:
\begin{equation}
\!\mathrm{SINR}_O = \frac{|h_1|^2 \, \Delta_1 \,\ell(r_1)}{\sigma + \mathrm{I}}, \,\,\,\,\,\text{where}\,\,\, \mathrm{I} = \sum_{j = 2}^b |h_j|^2 \,\Delta_j\, \ell(r_j).\!\!\!\!\label{eq.sinr_O}
\end{equation}
Terms $h_i$ and $\Delta_i$ are the small-scale fading component and the overall transmit/receive antenna gain associated with BS $i$, respectively, for $i = 1, \ldots, b$. The term $\mathrm{I}$ is the total interference contribution determined by all the BSs except the one connected to the standard user, i.e., the total interference determined by BSs $2, \ldots, b$. From Assumption~\ref{ass.BSst} and~\ref{ass.antennaPattern}, it follows that $\Delta_1$ is equal to $G_{\mathrm{TX}}G_{\mathrm{RX}}$. Finally, $\sigma$ represents the thermal noise power normalized with respect to the transmission power $P_t$.

As acknowledged in~\cite{RR0} there is a lack of extensive measurements for vehicular mmWave networks, as well as widely accepted channel models. Therefore, it is necessary to adopt conservative assumptions on signal propagation.
As summarized in Table~\ref{tab.refs}, several channel models have been proposed in the literature. 
Typically publicly available system-level mmWave simulators~\cite{ns3mmw} adopt channel models entirely~\cite{6932503} or partially~\cite{7370940} based on the Nakagami model, which are more refined alternatives to the widely adopted models dictating constant small-scale contributions~\cite{C1,C2,7105406}. In particular, we adopt the same channel model in~\cite{7370940}, which is based on the following observations: (i) because of the beamforming capabilities and the sectorial antenna pattern, the signal is impaired only by a limited number of scatterers, and (ii) the interfering transmissions cluster with many scatterers and reach the standard user. Furthermore, the considered sectorial antennal model at the transmitter and receiver sides (see Assumption~\ref{ass.Pattern}) significantly reduces the angular spread of the incoming signals -- thus reducing the Doppler spread. Moreover, the incoming signals are concentrated in one direction. Hence, it is likely that there is a non-zero bias in the Doppler spectrum, which can be compensated by the automatic frequency control loop at the receiver side~\cite{5783993}. For these reasons, the Doppler effect has been assumed to be mitigated.
\begin{assumption}[Channel Model]\label{ass.ch}
The channel between the serving BS and standard user is described by a Nakagami channel model with parameter $m$, and hence, $|h_1|^2$ follows a gamma distribution (with shape parameter $m$ and rate equal to $1$). On the other hand, to capture the clustering of interfering transmissions the channels between the standard user and each interfering BS are modelled as independent Rayleigh channels -- thus $|h_2|^2, \ldots, |h_b|^2$ are independently and identically distributed as an exponential distribution with mean equal to $1$.
\end{assumption}

\subsection{Analytical Characterization of $\mathrm{I}$}\label{subsec.anI}
In order to provide an analytical characterization of the interference power at $O$, it is convenient to split the term $\mathrm{I}$ into four different components: (i) $\mathrm{I}_{\mathrm{U,L}}$ and $\mathrm{I}_{\mathrm{U,N}}$ representing the interference power associated with LOS and NLOS BSs placed on the upper side of the road whose positions are defined by the PPPs $\Phi_{\mathrm{U},\mathrm{L}}$ and $\Phi_{\mathrm{U},\mathrm{N}}$, respectively, and, (ii) $\mathrm{I}_{\mathrm{B,L}}$ and $\mathrm{I}_{\mathrm{B,N}}$ the interference power generated by LOS and NLOS BSs on the bottom side of the road placed at the location given by the PPPs $\Phi_{\mathrm{B},\mathrm{L}}$ and $\Phi_{\mathrm{B},\mathrm{N}}$. Overall, the total interference power is given by $\mathrm{I} = \sum_{\mathrm{S} \in \{\mathrm{U},\mathrm{B}\}, \mathrm{E} \in \{\mathrm{L},\mathrm{N}\}} \mathrm{I}_{\mathrm{S},\mathrm{E}}$. In addition, the relations $\Phi_{\mathrm{L}} = \Phi_{\mathrm{U},\mathrm{L}} \bigcup \Phi_{\mathrm{B},\mathrm{L}}$ and $\Phi_{\mathrm{N}} = \Phi_{\mathrm{U},\mathrm{N}} \bigcup \Phi_{\mathrm{B},\mathrm{N}}$ hold.

In the following result, we derive an approximation for the Laplace transform $\mathcal{L}_{\mathrm{I}}(s)$ of $\mathrm{I}$.

\begin{table}[]
\centering
\caption{Values of $(a,b,\Delta)$ for different $<\mathrm{|x_1|,\mathbb{S}_1,\mathbb{E}_1,S,E}>$.}
\label{tab.1}
{\scriptsize\begin{tabular}{|c|c|c|}
\hline
\textbf{Configuration of}                                                              & \textbf{Conditions on} $|x_1|$                                                                                & \textbf{Enumeration of elements}\\
$\hspace{-2mm}\mathrm{<\mathbb{S}_1,\mathbb{E}_1,S,E>}\hspace{-2mm}$                                                              &                                                                                 & $\hspace{-2mm}(a,b,\Delta) \in \mathcal{C}_{\mathrm{|x_1|,\mathbb{S}_1,\mathbb{E}_1,S,E}}\hspace{-2mm}$                                                                                                                                                                                                   \\ \hline\hline
\multirow{2}{*}{$<\mathrm{U,L,U,L}>$}& \begin{tabular}[c]{@{}c@{}}For any $|x_1|$\\ such that $J > 0$ \end{tabular}               & \begin{tabular}[c]{@{}c@{}}$(|x_1|,K,g_\mathrm{TX}G_\mathrm{RX})$,\\ $(K,+\infty,g_\mathrm{TX}g_\mathrm{RX})$,\\ $(|x_1|,+\infty,g_\mathrm{TX}g_\mathrm{RX})$\end{tabular}                                          \\ \cline{2-3} 
                                                                                & \begin{tabular}[c]{@{}c@{}}For any $|x_1|$\\ such that $J \leq 0$\end{tabular} & \begin{tabular}[c]{@{}c@{}}$(|x_1|,K,g_\mathrm{TX}G_\mathrm{RX})$,\\ $(K,+\infty,g_\mathrm{TX}g_\mathrm{RX})$,\\ $(|x_1|,|J|,g_\mathrm{TX}G_\mathrm{RX})$,\\ $(|J|,+\infty,g_\mathrm{TX}g_\mathrm{RX})$\end{tabular} \\ \hline
\multirow{2}{*}{$<\mathrm{U,L,U,N}>$}& \begin{tabular}[c]{@{}c@{}}For any $|x_1|$\\ such that $J > 0$\end{tabular}               & \begin{tabular}[c]{@{}c@{}}$(x_\mathrm{N}(r_1),J,g_\mathrm{TX}g_\mathrm{RX})$,\\ $(x_\mathrm{N}(r_1),+\infty,g_\mathrm{TX}g_\mathrm{RX})$,\\ $(J,K,g_\mathrm{TX}G_\mathrm{RX})$,\\ $(K,+\infty,g_\mathrm{TX}g_\mathrm{RX})$\end{tabular}                                          \\ \cline{2-3} 
 & \begin{tabular}[c]{@{}c@{}}For any $|x_1|$\\ such that $J \leq 0$\end{tabular} & \begin{tabular}[c]{@{}c@{}}Refer to the case\\ $\mathrm{<U,L,U,L>}$ ($J \leq 0$)\\ and replace $|x_1|$\\ with $x_\mathrm{N}(r_1)$\end{tabular} \\ \hline
$\mathrm{<U,L,B,L>}$                                                              & For any $|x_1|$                                                                                                & \begin{tabular}[c]{@{}c@{}}$(|x_1|,+\infty,g_\mathrm{TX}g_\mathrm{RX})$,\\ $(|x_1|,+\infty,g_\mathrm{TX}g_\mathrm{RX})$,\end{tabular}                                                                                \\ \hline
$\mathrm{<U,L,B,N>}$                                                              & \multicolumn{2}{c|}{\begin{tabular}[c]{@{}c@{}}Refer to the case $\mathrm{<U,L,B,L>}$ and\\ replace $|x_1|$ with $x_\mathrm{N}(r_1)$\end{tabular}}                                                                                                                                                                    \\ \hline
\multirow{2}{*}{$<\mathrm{U,N,U,L}>$}& \begin{tabular}[c]{@{}c@{}}For any $|x_1|$\\ \hspace{-1mm}such that $x_\mathrm{L}(r_1) > K\hspace{-1mm}$\end{tabular}               & \begin{tabular}[c]{@{}c@{}}Refer to the case\\ $\mathrm{<U,L,B,L>}$ and\\ replace $|x_1|$ with $x_\mathrm{L}(r_1)$\end{tabular}                                          \\ \cline{2-3}
& \begin{tabular}[c]{@{}c@{}}For any $|x_1|$\\ \hspace{-1mm}such that $x_\mathrm{L}(r_1) \leq K\hspace{-1mm}$\end{tabular} & \begin{tabular}[c]{@{}c@{}}Refer to the case\\ $\mathrm{<U,L,U,L>}$ and\\ replace $|x_1|$ with $x_\mathrm{L}(r_1)$\end{tabular} \\ \hline
$\mathrm{<U,N,U,N>}$                                                              & \multicolumn{2}{c|}{Refer to the case $\mathrm{<U,L,U,L>}$}                                                                                                                                                                                                                                                             \\ \hline
$\mathrm{<U,N,B,L>}$                                                              & \multicolumn{2}{c|}{\begin{tabular}[c]{@{}c@{}}Refer to the case $\mathrm{<U,L,B,L>}$ and\\ replace $x_1$ with $x_\mathrm{L}(r_1)$\end{tabular}}                                                                                                                                                          \\ \hline
$\mathrm{<U,N,B,N>}$                                                              & \multicolumn{2}{c|}{Refer to the case $\mathrm{<U,L,B,L>}$}                                                                                                                                                                                                                                                             \\ \hline
\begin{tabular}[c]{@{}c@{}}Cases where\\ $\mathbb{S}_1 = \mathrm{B}$, $\mathrm{S} = \mathrm{B}$\end{tabular} & \multicolumn{2}{c|}{\begin{tabular}[c]{@{}c@{}}Refer to the correspondent cases \\ where $\mathbb{S}_1 = \mathrm{U}$ and $\mathrm{S} = \mathrm{U}$\end{tabular}}                                                                                                                                                                                      \\ \hline
\begin{tabular}[c]{@{}c@{}}Cases where\\ $\mathbb{S}_1 = \mathrm{B}$, $\mathrm{S} = \mathrm{U}$\end{tabular} & \multicolumn{2}{c|}{\begin{tabular}[c]{@{}c@{}}Refer to the correspondent cases \\ where $\mathbb{S}_1 = \mathrm{U}$ and $\mathrm{S} = \mathrm{B}$\end{tabular}}                                                                                                                                                                                      \\ \hline
\end{tabular}}
\end{table}

\begin{theorem}\label{th.LI}
Let $\mathbb{S}_1 = \mathrm{U}$  and $\mathbb{S}_1 = \mathrm{B}$ represent the cases where the standard user connects to a BS on the upper or the bottom side of the road, respectively. In addition, let $\mathbb{E}_1 = \mathrm{L}$ and $\mathbb{E}_1 = \mathrm{N}$ signify the cases where the standard user connects to a LOS or NLOS BS, respectively.
The Laplace transform $\mathcal{L}_{\mathrm{I}_{\mathrm{S,E}},\mathbb{E}_1}(s)$ of $\mathrm{I}_{\mathrm{S,E}}$, conditioned on $\mathbb{E}_1$, for $\mathrm{S} \in \{\mathrm{U},\mathrm{B}\}$ and $\mathrm{E} \in \{\mathrm{L},\mathrm{N}\}$, can be approximated as follows:
	\begin{equation}
	\mathcal{L}_{\mathrm{I}_{\mathrm{S,E}},\mathbb{E}_1}(s) \cong \hspace{-6mm}\mathop{\mathop{\mathop{\prod_{\mathbb{S}_1\in\{\mathrm{U,B}\},}}}_{(a,b,\Delta) \in \mathcal{C}_{\mathrm{|x_1|,\mathbb{S}_1,\mathbb{E}_1,S,E}}}} \hspace{-6mm} \sqrt{\mathcal{L}_{\mathrm{I}_{\mathrm{S,E}},\mathbb{E}_1}(s;a,b,\Delta)},\label{eq.Th1}
	\end{equation}
where $\mathcal{L}_{\mathrm{I}_{\mathrm{S,E}},\mathbb{E}_1}(s;a,b,\Delta)$ is defined as in~\eqref{eq.app.pTh1}.
We define the $x$-axis coordinates $J = w(N_o + 1) / [\tan(\epsilon^{\mathrm{(U)}} + \psi/2)]$ and $K = w(N_o + 1) / [\tan(\epsilon^{\mathrm{(U)}} - \psi/2)]$ of the points where the two rays defining the standard user beam intersect with a side of the road, where \mbox{$\epsilon^{\mathrm{(U)}} = \mp\tan^{-1}[w(N_o + 1)/x_1]$}, for $\mathbb{S}_1 = \mathrm{U}$ or $\mathrm{B}$, respectively (see Fig.~\ref{fig.rxGainExpl}). Furthermore, let us define $x_\mathrm{L}(r_1) = \sqrt{(\mathrm{A}_\mathrm{L}(r_1))^2 - w^2(N_o+1)^2}$ and \mbox{$x_\mathrm{N}(r_1) = \sqrt{(\mathrm{A}_\mathrm{N}(r_1))^2 - w^2(N_o+1)^2}$}. Different combinations of parameters $<\mathrm{|x_1|,\mathbb{S}_1,\mathbb{E}_1,S,E}>$ determine different sequences $\mathcal{C}_{\mathrm{|x_1|,\mathbb{S}_1,\mathbb{E}_1,S,E}}$ of parameter configurations $(a,b,\Delta)$, as defined in Table~\ref{tab.1}.
\end{theorem}
\begin{IEEEproof}
See Appendix~\ref{app.A}.
\end{IEEEproof}

\begin{example}
Consider the scenario where the standard user connects to a LOS BS, i.e., $\mathbb{E}_1 = \mathrm{L}$, and relation $J > 0$ holds.
We evaluate the Laplace transform of the interference associated with the BSs located on the upper side of the road ($\mathrm{S} = \mathrm{U}$) that are in LOS with respect to the standard user ($\mathrm{E} = \mathrm{L}$). The sequence $\mathcal{C}_{\mathrm{|x_1|,U,L,U,L}}$ is given by the first row of Table~\ref{tab.1}, while $\mathcal{C}_{\mathrm{|x_1|,B,L,U,L}}$ consists of the same elements of sequence $\mathcal{C}_{\mathrm{|x_1|,U,L,B,L}}$ (last row of Table~\ref{tab.1}). As a result, $\mathcal{L}_{\mathrm{I}_{\mathrm{S,E}},\mathbb{E}_1}(s)$ can be approximated as follows:
\setlength{\arraycolsep}{0.0em} 
\begin{eqnarray}
\mathcal{L}_{\mathrm{I}_{\mathrm{S,E}},\mathbb{E}_1}(s) &{}\cong{}& \Big[{\mathcal{L}_{\mathrm{I}_{\mathrm{S,E}},\mathbb{E}_1}(s;|x_1|,K,g_\mathrm{TX}G_\mathrm{RX})}\notag\\
&& \cdot \, {\mathcal{L}_{\mathrm{I}_{\mathrm{S,E}},\mathbb{E}_1}(s;K,+\infty,g_\mathrm{TX}g_\mathrm{RX})}\nonumber\\
&& \cdot \, {\mathcal{L}_{\mathrm{I}_{\mathrm{S,E}},\mathbb{E}_1}(s;|x_1|,+\infty,g_\mathrm{TX}g_\mathrm{RX})}\Big]^{1/2}\nonumber\\
&& \cdot \, {\mathcal{L}_{\mathrm{I}_{\mathrm{S,E}},\mathbb{E}_1}(s;x_\mathrm{N}(r_1),+\infty,g_\mathrm{TX}g_\mathrm{RX})}.
\end{eqnarray}
\end{example}

From Theorem~\ref{th.LI} and the fact that $\mathrm{I}$ is defined as a sum of statistically independent interference components, the following corollary holds.

\begin{corollary}\label{cor.lI}
The Laplace transform of $\mathrm{I}$, for $\mathbb{E}_1 = \{\mathrm{L},\mathrm{N}\}$, can be approximated as follows:
\begin{equation}
\mathcal{L}_\mathrm{I,\mathbb{E}_1}(s) \cong \prod_{\mathrm{S} \in \{\mathrm{U},\mathrm{B}\}, \mathrm{E} \in \{\mathrm{L},\mathrm{N}\}} \mathcal{L}_{I_{\mathrm{S},\mathrm{E}},\mathbb{E}_1}(s)
\end{equation}
\end{corollary}

\begin{example}
Consider the scenario where $\mathbb{E}_1 = \mathrm{L}$, and relation $J > 0$ holds. Using Corollary~\ref{cor.lI}, $\mathcal{L}_{\mathrm{I}_{\mathrm{S,E}},\mathbb{E}_1}(s)$ can be approximated as follows:
\setlength{\arraycolsep}{0.0em} 
\begin{eqnarray}
\mathcal{L}_{\mathrm{I},\mathbb{E}_1}(s) &{}\cong{}& {\mathcal{L}_{\mathrm{I}_{\mathrm{S,E}},\mathbb{E}_1}(s;|x_1|,K,g_\mathrm{TX}G_\mathrm{RX})}\notag\\
&& \cdot \, {\mathcal{L}_{\mathrm{I}_{\mathrm{S,E}},\mathbb{E}_1}(s;x_\mathrm{N}(r_1),J,g_\mathrm{TX}g_\mathrm{RX})}\nonumber\\
&& \cdot \, {\mathcal{L}_{\mathrm{I}_{\mathrm{S,E}},\mathbb{E}_1}(s;J,K,g_\mathrm{TX}G_\mathrm{RX})}\nonumber\\
&& \cdot \, \left({\mathcal{L}_{\mathrm{I}_{\mathrm{S,E}},\mathbb{E}_1}(s;K,+\infty,g_\mathrm{TX}g_\mathrm{RX})}\right)^2\nonumber\\
&& \cdot \, \left({\mathcal{L}_{\mathrm{I}_{\mathrm{S,E}},\mathbb{E}_1}(s;|x_1|,+\infty,g_\mathrm{TX}g_\mathrm{RX})}\right)^3\nonumber\\
&& \cdot \, \left({\mathcal{L}_{\mathrm{I}_{\mathrm{S,E}},\mathbb{E}_1}(s;x_\mathrm{N}(r_1),+\infty,g_\mathrm{TX}g_\mathrm{RX})}\right)^3
\end{eqnarray}
\end{example}

\subsection{SINR Outage and Rate Coverage Probability Framework}
The general framework for evaluating the SINR outage probability is given in the following result.
\begin{theorem}\label{th.SINRout}
Let
\begin{equation}
\mathrm{F}_\mathrm{L}(t) = e^{-2 \lambda_\mathrm{L} \sqrt{t^2-w^2(N_o+1)^2}}
\end{equation}
and
\begin{equation}
\mathrm{F}_\mathrm{N}(t) = e^{-2 \lambda_\mathrm{N} \sqrt{t^2-w^2(N_o+1)^2}}\label{eq.FN}
\end{equation}
be the probability of a LOS or NLOS BS not being at a distance smaller than $t$ from $O$, respectively. We regard $\mathrm{P}_\mathrm{T}(\theta)$ to be the SINR outage probability with respect to a threshold $\theta$, i.e., the probability that $\mathrm{SINR}_O$ is smaller than a threshold $\theta$. $\mathrm{P}_\mathrm{T}(\theta)$ can be expressed as follows:
\setlength{\arraycolsep}{0.0em} 
\begin{eqnarray}
\mathrm{P}_\mathrm{T}(\theta) &{}={}& \mathrm{P}_{\mathrm{L}} - \overbrace{\mathbb{P}[\textrm{SINR}_O > \theta \,\,\text{$\land$}\,\, \text{std. user served in LOS}]}^{\mathrm{P}_\mathrm{CL}(\theta)}\nonumber\\
 &&\hspace{-13mm} +\,\mathrm{P}_{\mathrm{N}} - \underbrace{\mathbb{P}[\textrm{SINR}_O > \theta \,\,\text{$\land$}\,\, \text{std. user served in NLOS}]}_{\mathrm{P}_\mathrm{CN}(\theta)}\label{eq.sinrOut.Th}
\end{eqnarray}
where
\setlength{\arraycolsep}{0.0em}
\begin{eqnarray}
\mathrm{P}_\mathrm{CL}(\theta) &{}=_{\Big|_{\mathbb{E}_1 = \mathrm{L}}}{}& \hspace*{-4mm}-\sum_{k = 0}^{m-1}(-1)^{m-k}\binom{m}{k}\int_{w(N_o+1)}^{+\infty} e^{-\frac{v\sigma\theta(m-k)}{\Delta_1 \mathrm{C}_\mathrm{L}}r_1^{\alpha_\mathrm{L}}}\notag\\
&& \hspace{-13mm}\cdot \, \mathcal{L}_\mathrm{I,\mathbb{E}_1}\left(\frac{v\theta r_1^{\alpha_{\mathrm{L}}}(m-k)}{\Delta_1 \mathrm{C}_\mathrm{L}}\right) f_\mathrm{L}(r_1) \mathrm{F}_\mathrm{N}(\mathrm{A}_\mathrm{N}(r_1)) \, d r_1\label{eq.PCL}
\end{eqnarray}
and
\setlength{\arraycolsep}{0.0em} 
\begin{eqnarray}
\mathrm{P}_\mathrm{CN}(\theta) &{}=_{\Big|_{\mathbb{E}_1 = \mathrm{N}}}{}& \hspace*{-4mm}-\sum_{k = 0}^{m-1}(-1)^{m-k}\binom{m}{k}\int_{w(N_o+1)}^{+\infty} e^{-\frac{v\sigma\theta(m-k)}{\Delta_1 \mathrm{C}_\mathrm{N}}r_1^{\alpha_\mathrm{N}}}\notag\\
&& \hspace{-13mm}\cdot \, \mathcal{L}_\mathrm{I,\mathbb{E}_1}\left(\frac{v\theta r_1^{\alpha_{\mathrm{N}}}(m-k)}{\Delta_1 \mathrm{C}_\mathrm{N}}\right) f_\mathrm{N}(r_1) \mathrm{F}_\mathrm{L}(\mathrm{A}_\mathrm{L}(r_1)) \, d r_1,\label{eq.PCN}
\end{eqnarray}
represent the probability of the standard user not experiencing SINR outage while connected to a LOS or NLOS BS, respectively.
\end{theorem}
\begin{IEEEproof}
The result~\eqref{eq.sinrOut.Th} follows immediately once $\mathrm{P}_\mathrm{CL}(\theta)$ and $\mathrm{P}_\mathrm{CN}(\theta)$ as known. In particular, the following relation holds (for $\mathbb{E}_1 = \mathrm{L}$):
\setlength{\arraycolsep}{0.0em} 
\begin{eqnarray}
\mathrm{P}_\mathrm{CL}(\theta) &{}={}& \mathbb{P}\Bigg[\frac{|h_1|^2 \,\Delta_1\, \ell(r_1)}{\sigma + \mathrm{I}} > \theta \,\,\text{$\land$ std. user served in LOS}\Bigg]\notag\\
&{}\stackrel{(i)}{\cong}{}& \mathbb{E}_\mathrm{I} \int_{w(N_o+1)}^{+\infty}      \left( 1 - \left(1-e^{-v\frac{(\sigma + \mathrm{I})\theta}{\Delta_1 \mathrm{C}_\mathrm{L}}r_1^{\alpha_\mathrm{L}}}\right)^m\right)\notag\\
&& \cdot \,f_\mathrm{L}(r_1) \mathrm{F}_\mathrm{N}(\mathrm{A}_\mathrm{N}(r_1)) \, d r_1
\end{eqnarray}
where $v = m(m!)^{-1/m}$~\cite[Lemma 6]{6932503} and $(i)$ arise from $|h_1|^2$ being distributed as a gamma random variable. In addition, $\mathrm{F}_\mathrm{N}(\mathrm{A}_\mathrm{N}(r_1))$ is defined as the probability of a NLOS BS not being at a distance smaller than $\mathrm{A}_\mathrm{N}(r_1)$ to $O$, i.e., the probability that the standard user is not connected to a NLOS BS. The expression of $\mathrm{F}_\mathrm{N}(t)$, as in~\eqref{eq.FN}, immediately follows from the simplification of the following relation:
\begin{equation}
\mathrm{F}_\mathrm{N}(t) = 1 - \int_{w(N_o+1)}^{t} f_\mathrm{N}(r) \, dr.
\end{equation}
From the binomial theorem, we swap the integral and the expectation with respect to $\mathrm{I}$ and invoke Corollary~\ref{cor.lI} to obtain~\eqref{eq.PCL}. By following the same reasoning, it is also possible to derive expressions for $\mathrm{P}_\mathrm{CN}(\theta)$ and $\mathrm{F}_\mathrm{L}(t)$.
\end{IEEEproof}

{\begin{remark}\label{r.app.gen}
As the value of $\alpha_{\mathrm{N}}$ increases it is less likely that the standard user connects to a NLOS BS. Hence, from~\eqref{eq.An}, $A_{\mathrm{N}}$ is likely to be equal to $w(N_o + 1)$. As a result, the exponential term in~\eqref{eq.Pl} approaches one and, hence, $\mathrm{P}_{\mathrm{L}}$ can be approximated as follows:
\begin{equation}
\mathrm{P}_{\mathrm{L}} \cong \int_{w(N_o+1)}^{\infty} f_\mathrm{L}(r) \,dr = \int_{0}^{\infty} 2 \lambda_{\mathrm{L}} e^{-2 \lambda_{\mathrm{L}} t} = 1.
\end{equation}
Using this approximation, it follows that \mbox{$\mathrm{P}_{\mathrm{N}} \cong 0$} holds. In addition, since $\mathrm{P}_{\mathrm{CN}}$ is always less than or equal to $\mathrm{P}_{\mathrm{N}}$, the relation $\mathrm{P}_\mathrm{CN} \cong 0$ holds as well. If \mbox{$A_{\mathrm{N}}\cong w(N_o + 1)$}, the relation $\mathrm{F}_\mathrm{N}(\mathrm{A}_\mathrm{N}(r_1))\cong 1$ holds.
For these reasons, $\mathrm{P}_{\mathrm{T}}(\theta)$ can be approximated as follows:
\setlength{\arraycolsep}{0.0em}
\begin{eqnarray}
\mathrm{P}_\mathrm{T}(\theta) &{}\cong{}& 1 + \sum_{k = 0}^{m-1}(-1)^{m-k}\binom{m}{k}\int_{w(N_o+1)}^{+\infty} \!\!\!\!\!\!\!\!\!\!\!\!e^{-\frac{v\sigma\theta(m-k)}{\Delta_1 \mathrm{C}_\mathrm{L}}r_1^{\alpha_\mathrm{L}}}\notag\\
&& \cdot \, \mathcal{L}_\mathrm{I,\mathrm{L}}\left(\frac{v\theta r_1^{\alpha_{\mathrm{L}}}(m-k)}{\Delta_1 \mathrm{C}_\mathrm{L}}\right) f_\mathrm{L}(r_1)  \, d r_1. \label{eq.simplPT}
\end{eqnarray}
In addition, should signals from NLOS BSs be entirely attenuated by blockages, $\mathrm{P}_{\mathrm{N}}$ would be equal to $0$ and~\eqref{eq.simplPT} would hold as well.
\end{remark}}

{\begin{remark}
Should the BSs be deployed only along the horizontal line separating the two driving directions, the value of $p_\mathrm{L}$ be equal to $1$ and, hence, $P_\mathrm{L} = 1$. Under this circumstances \eqref{eq.simplPT} would hold with minimal changes. For instance, let us focus on the East-to-West driving direction, assume that the standard user be located in the middle of the $N_o + 1$ lanes, and that the origin of the coordinate system overlaps with the standard user position. In this case, an interfering BS $i$ is associated with $\epsilon_i$, which takes values uniformly distributed in $[0, 2\pi)$. Let us regard with $\mathrm{U}$ the upper-most edge of the East-to-West driving direction, $\mathbb{E}_1$ is equal to $\mathrm{L}$ and hence, relation $\mathcal{L}_\mathrm{I,\mathrm{L}}(s) \cong \mathcal{L}_{I_{\mathrm{U},\mathrm{L}},\mathrm{L}}(s)$ holds. This allows us to approximate $\mathrm{P_T}$ as in~\eqref{eq.simplPT}, where term $w(N_o+1)$ has to be replaced with $w(N_o+1)/2$.
\end{remark}}

From~\cite[Theorem 1]{6497002} and by using Theorem~\ref{th.SINRout}, it is now possible to express the rate coverage probability $\mathrm{R}_\mathrm{C}(\kappa)$, i.e., the probability that the standard user experiences a rate that is greater than or equal to $\kappa$. In particular, the rate coverage probability is given by:
\setlength{\arraycolsep}{0.0em} 
\begin{eqnarray}
\mathrm{R}_\mathrm{C}(\kappa) &{}={}& \mathbb{P}[\text{rate of std. user $\geq$ $\kappa$}]\notag\\
&{}={}& 1 - \mathrm{P}_\mathrm{T}(2^{\kappa/W}-1),\label{eq.rate}
\end{eqnarray}
where $W$ is the system bandwidth.

\begin{table}[t]
\centering
\caption{Main simulation parameters.}
\label{tab.sim}
{\scriptsize
\begin{tabular}{|P{2cm}|P{5.8cm}|}
\hline \textbf{Parameter}  & \textbf{Value}\\
\hline \vspace{-2mm}Simulated time & \vspace{-2mm}\SI{13}{\hour} (for Figs.~\ref{fig.f_1}-\ref{fig.f_4_SIM}), \SI{55}{\hour} (for Fig.~\ref{fig.f_0})\\
\hline Length of the simulated road section (i.e., $2R$) & \vspace{1mm}\SI{20}{\kilo\meter}, \SI{100}{\kilo\meter}\\
\hline $w$ & \SI{3.7}{\meter}, as per~\cite{ioannou2013automated}\\
\hline \vspace{-1.6mm}$\lambda_u$ & \vspace{-1.6mm}$2 \cdot 10^{-2}$\\
\hline \vspace{-2mm}Mobility model & \vspace{-2mm}Blockages and the standard user move according to a Krauss car-following mobility model~\cite{Kanagaraj2013390}; maximum acceleration and deceleration equal to \SI{5.3}{\meter\per\second^2}~\cite{fambro1997determination}, maximum vehicle speed equal to \SI{96}{\kilo\meter\per\hour} (blockages) and \SI{112}{\kilo\meter\per\hour} (standard user). \\
\hline \vspace{-2mm}Blockage dimensions & \vspace{-2mm}The dimensions of a double decker bus, i.e.,  length $\tau$ equal to \SI{11.2}{\meter} and width equal to \SI{2.52}{\meter}~\cite{routermaster}\\
\hline $N_o$ & $1$, $2$\\
\hline $\{\lambda_{o,1},\lambda_{o,2}\}$ & $\{1 \cdot 10^{-2}, 2 \cdot 10^{-2}\}$, i.e., one blockage every $\{\SI{100}{\meter}, \SI{50}{\meter}\}$\\
\hline $\lambda_\mathrm{BS}$ & From $2 \cdot 10^{-4}$ to $1 \cdot 10^{-2}$, with a step of $2 \cdot 10^{-4}$\\
\hline \vspace{-1.6mm}Carrier frequency $f$ & \vspace{-1.6mm}$\SI{28}{\giga\hertz}$\\
\hline \vspace{-1.6mm}$\mathrm{C_L}$, $\mathrm{C_N}$ & \vspace{-1.6mm}$-20 \log_{10}(4 \pi f / c)$, which is the free space path loss in dB at a distance of \SI{1}{\meter} and $c$ is the speed of light~\cite{7481506}\\
\hline $\alpha_\mathrm{L}$ & $2.8$\\
\hline \vspace{-1.6mm}$\alpha_\mathrm{N}$ & \vspace{-1.6mm}$\{4,5.76\}$, as per~\cite{6387266}\\
\hline $m$ & $3$, as per~\cite{7370940}\\
\hline $\phi$ & $\{30^\circ,90^\circ\}$\\
\hline $G_\mathrm{TX}$ & $\{\SI{10}{\dB},\SI{20}{\dB}\}$\\
\hline $G_\mathrm{RX}$ & \SI{10}{\dB}\\
\hline $g_\mathrm{TX}$, $g_\mathrm{RX}$ & \SI{-10}{\dB}\\
\hline $W$ & \SI{100}{\mega\hertz}\\
\hline $P_t$ & \SI{27}{\dBm}\\
\hline Thermal noise power (i.e, $\sigma \cdot P_t$) & \mbox{$10 \log_{10}(k \cdot T \cdot W \cdot 10^3)$ \SI{}{\dBm}}, where $k$ is the Boltzmann constant and the temperature $T = \SI{290}{\kelvin}$~\cite{rappaport2014millimeter}\\
\hline
\end{tabular}
}
\end{table}

\section{Numerical Results}\label{sec.nr}
\subsection{Simulation Framework}\label{subsec.sc}
{In order to validate the proposed theoretical model, we developed a novel MATLAB simulation framework capable of estimating the SINR outage and rate coverage probabilities by means of the Monte Carlo approach. Both our simulator and the implementation of the proposed theoretical framework are available online~\cite{SimFW}.}

\begin{figure}[t]
\centering
\includegraphics[width=1\columnwidth]{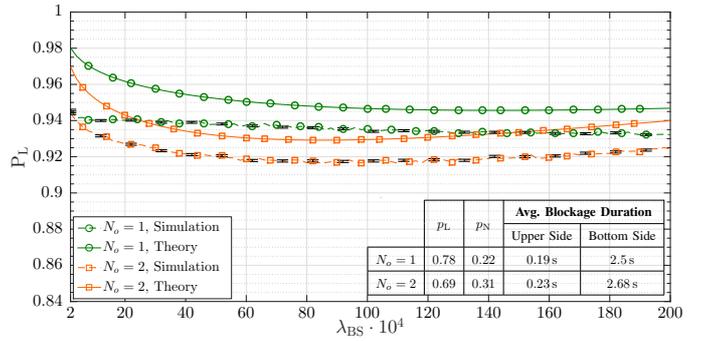}
\caption{Probability $\mathrm{P}_\mathrm{L}$ that the standard users connects to a LOS BS as a function of $\lambda_{\mathrm{BS}}$, for $N_o = \{1,2\}$ and $\alpha_\mathrm{N} = 4$.}
\label{fig.f_0}
\end{figure}

We remark that Assumption~\ref{ass.roadL} models the highway as infinitely long, which is not possible in a simulation. However, as noted in~\cite{NET-015,weber2005computational}, the radius $R$ of the simulated system (i.e., the length of the simulated road section $2R$) can be related to the simulation accuracy error $\varepsilon$, as in~\cite[Eq.~(3.5)]{NET-015}. In the case of a one-dimensional PPP, the radius is related to the simulation error by $R \geq \varepsilon^{-\frac{1}{\alpha_\mathrm{L} - 1}}$. {We superimpose a normal approximation of the binomial proportion confidence interval~\cite{ryan2007modern} to our simulation results defined as $\left[ \hat{p} - z \sqrt{{\hat{p}(1-\hat{p})/n}}; \hat{p} + z \sqrt{{\hat{p}(1-\hat{p})/n}} \right]$, where $\hat{p}$ is the simulated probability value, $n$ is the number of Monte Carlo iterations and $z$ is the $(1-0.5 \cdot e)$-th quantile, for $0 \leq e \leq 1$. In particular, $z$ is set equal to $0.99$, defining a confidence intervals of $98\%$ and $n$ is equal to $2 \cdot 10^5$ (for Fig.~\ref{fig.f_0}) or $5 \cdot 10^4$ (for Figs.~\ref{fig.f_1}-\ref{fig.f_4_SIM}).}

We simulated scenarios where the standard user drives on the lower-most user lane along with multiple other vehicles; in particular, we considered a number of vehicles equal to $\lfloor2R\lambda_u\rfloor$, where $\lambda_u$ is the vehicle density driving on each user lane. In addition, a number of blockages equal to $\lfloor 2R\lambda_{o,i}\rfloor$ are placed at random on each obstacle lane $i$, for $i = 1, \ldots, N_o$. During each simulated scenario, both the vehicles driving on the user lanes and blockages move according to the Krauss car-following mobility model~\cite{Kanagaraj2013390} and their maximum speed is set equal to \SI{70}{\mph} (i.e., \SI{112}{\kilo\meter\per\hour}) and \SI{60}{\mph} (i.e., \SI{96}{\kilo\meter\per\hour}), respectively as dictated by the current British speed limits\footnote{We refers to the Highway Code (\url{https://www.gov.uk/speed-limits}) valid for England, Scotland and Wales.}. In order to keep the density of the simulated blockages constant and hence allow a fair validation of the proposed theoretical framework, we adopted the Krauss car-following mobility model with the wrap-around policy. In particular, when a vehicle reaches the end of the simulated road section, it re-enters at the beginning.

BSs are positioned uniformly at random at both sides of the road.
The simulator estimates the SINR outage probability $\mathrm{P_T}$ and rate coverage probability $\mathrm{R_C}$ by averaging over the total simulation time and across a number of BS random locations and steering angle configurations (of the interfering BSs); number allowing to the simulated average performance metrics to converge to a stable value.
We remark that the adoption of highly directional antennas significantly reduces the angular spread of the incoming signals. As such, in the simulated scenarios, we assume the standard user is equipped with an automatic frequency control loop compensating for the Doppler effect~\cite{5783993}.
In addition, the simulated channel follows Assumption~\ref{ass.ch}.

With regards to Table~\ref{tab.sim}, we consider $N_o = \{1,2\}$ obstacle lanes per driving direction. For $N_o = 2$, we assume different traffic intensities by setting densities $\{\lambda_{\mathrm{o},1}, \lambda_{\mathrm{o},2}\}$ as per row six of Table~\ref{tab.sim}. Furthermore, we consider a typical highway lane width $w$~\cite{ioannou2013automated}.

In Section~\ref{sec.LN}, we approximated the probabilities $p_\mathrm{L}$ and $p_\mathrm{N}$ for a BS of being in LOS or NLOS with respect to the standard user, respectively. It is worth noting that approximation~\eqref{eq.losNlosApp} has been invoked only in the derivation of the proposed theoretical model. {In contrast, in the simulated scenarios a BS is in NLOS only if the ideal segment connecting the standard user and the BS intersects with one or more vehicles in the obstacle lanes and not just with their footprint segments.}

Communications between the standard user and the serving BS are impaired only by large vehicles (namely, trucks, double-decker buses, etc.) driving on the obstacle lanes. Specifically, we consider blockages with length ($\tau$) and width of a double-decker bus~\cite{routermaster}. Without loss of generality, both the proposed theoretical framework and our simulations consider bi-dimensional scenarios. Although it is always possible to deploy BSs having an antenna height sufficient to prevent the vehicles in the obstacle lanes to behave as blockages, it is not always feasible in practice. For instance, in a $4$-lane road section ($N_o = 2$) with $w = \SI{3.7}{\meter}$ where the standard user drives in the middle of the lower-most user lane and the user antenna height is $\SI{1.5}{\meter}$, the BS antenna height should be greater than $\SI{12.5}{\meter}$ to allow a blockage-free scenario, which is at least twice as much the antenna height in a typical LTE-A urban deployment~\cite{3gppAntenna}.
Therefore, we assume that the BS antenna height is \SI{5}{\meter}, which means that vehicles in the obstacle lanes always behave as blockages.
For this reason, we do not further consider the height of the vehicles in our study. All the remaining simulation parameters are summarized in Table~\ref{tab.sim}.

\subsection{Theoretical Model Assessment}\label{subsec.val}
In order to numerically study our mmWave highway network and assess the accuracy of our theoretical model, we first consider $\alpha_{N} = 4$ and a road section with a length $2R = \SI{100}{\kilo\meter}$, which ensures a simulation accuracy error of at least $10^{-7.2}$. 
{In addition, the adoption of a relatively small but realistic value of $\alpha_{N}$ makes more likely for the standard user to a connect to an NLOS BS~\cite{6932503} and hence, allows us to effectively validate the proposed LOS/NLOS user association model (see Lemma~\ref{lem.serv}).}
Considering the density $\lambda_\mathrm{BS}$ of process $\Phi_\mathrm{BS}$, we ideally project the BSs onto the $x$-axis and we define their projected mean Inter-Site-Distance ($x$-ISD) as $1/\lambda_\mathrm{BS}$.

\begin{figure}[tb]
\centering
\subfloat[$\lambda_{\mathrm{BS}} = 10^{-2}$, $x$-ISD = $\SI{100}{\meter}$]{\label{fig.f_1.1}
	\includegraphics[width=1\columnwidth]{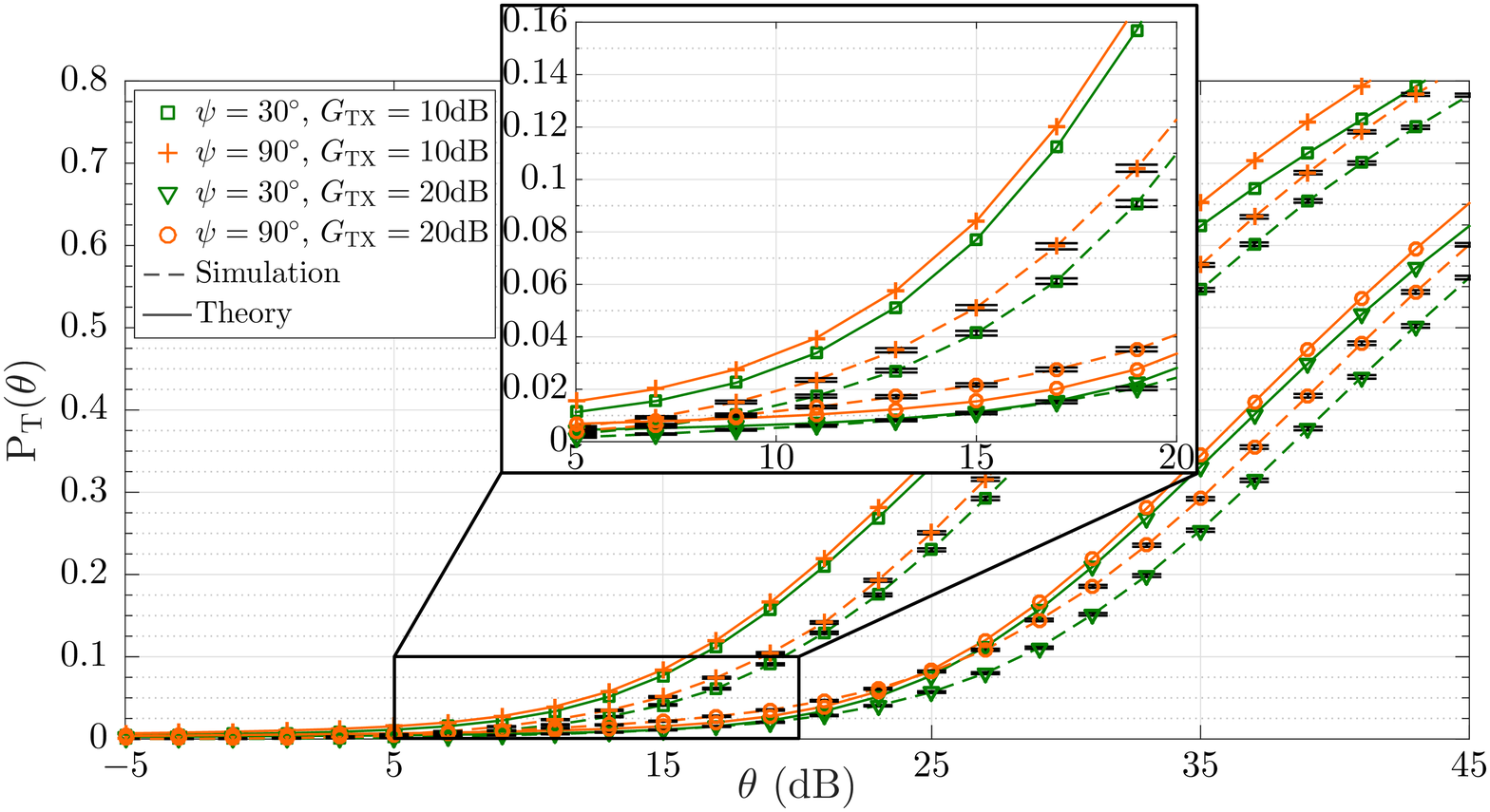}
}\\
\subfloat[$\lambda_{\mathrm{BS}} = 4 \cdot 10^{-3}$, $x$-ISD = $\SI{250}{\meter}$]{\label{fig.f_1.2}
	\includegraphics[width=1\columnwidth]{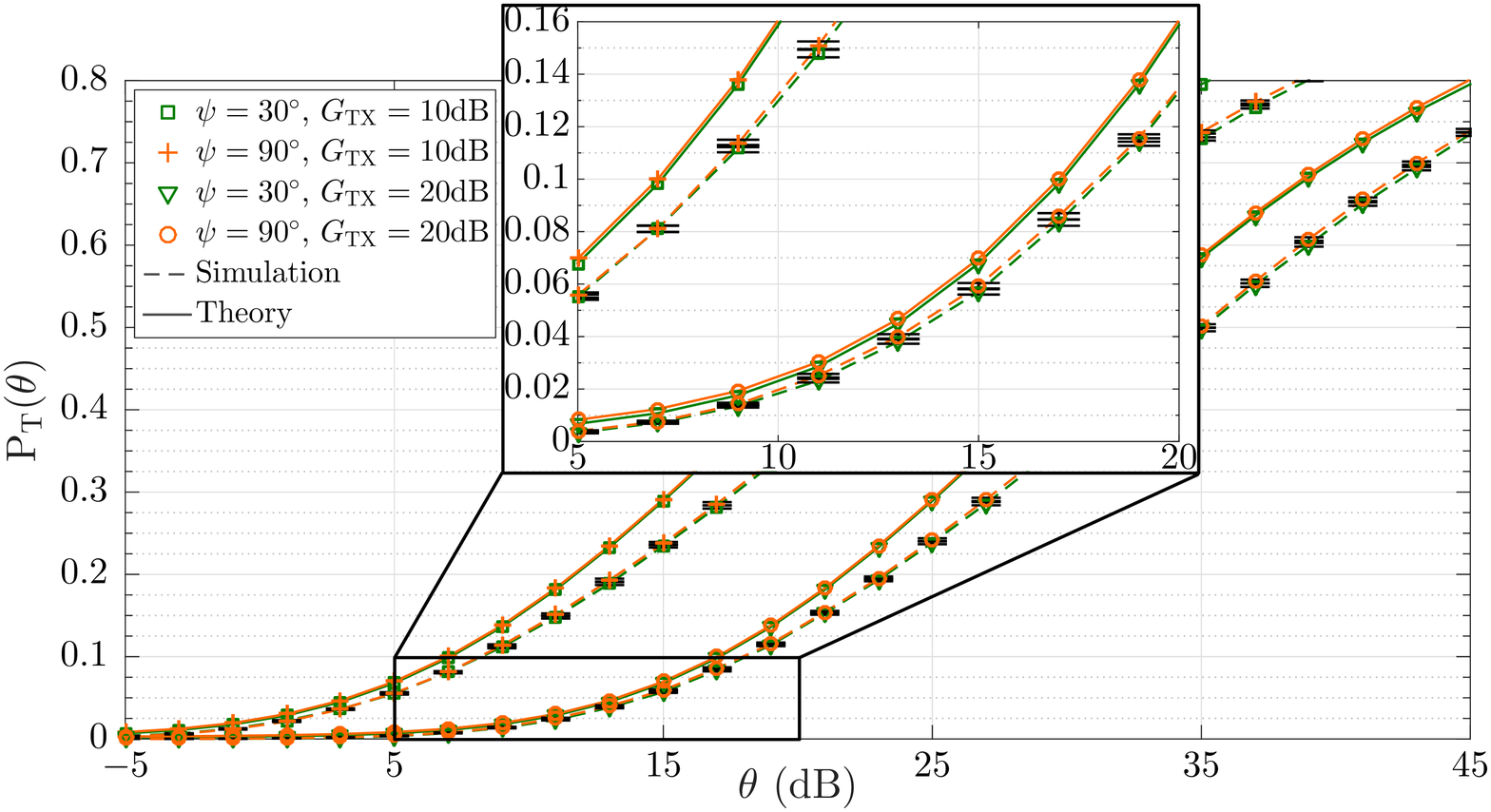}
}
\caption{SINR outage probability $\mathrm{P}_{\mathrm{T}}$ as a function of the threshold $\theta$, for $N_o = 1$, $\alpha_\mathrm{N} = 4$, $\psi = \{30^\circ, 90^\circ\}$ and $G_{\mathrm{TX}} = \{\SI{10}{\dB}, \SI{20}{\dB}\}$.}
\label{fig.f_1}
\end{figure}

Let us consider a Fig.~\ref{fig.f_0} shows the probability of the standard user connecting to a LOS BS as a function of $\lambda_{\mathrm{BS}}$ for one and two obstacle lanes on each side of the road.
The equivalent $x$-ISD spans between \SI{5}{\kilo\meter} ($\lambda_{\mathrm{BS}} = 2 \cdot 10^{-4}$) and \SI{50}{\meter} ($\lambda_{\mathrm{BS}} = 2 \cdot 10^{-2}$). In particular, as typically happens, we observe that $\mathrm{P}_\mathrm{L}$ is significantly greater than $\mathrm{P}_\mathrm{N}$. Specifically, if $N_o = 1$ then,  for $\lambda_{\mathrm{BS}} = 4 \cdot 10^{-3}$, the simulated value of $\mathrm{P}_\mathrm{L}$ is equal to $0.94$. When $N_o$ increases to $2$, the simulated value of $\mathrm{P}_\mathrm{L}$ reduces to $0.92$, for $\lambda_{\mathrm{BS}} = 4 \cdot 10^{-3}$.

Fig.~\ref{fig.f_0} also compares our approximated theoretical expression of  $\mathrm{P}_\mathrm{L}$, as in~\eqref{eq.Pl}, with the simulated one. We note that~\eqref{eq.Pl} overestimates $\mathrm{P}_\mathrm{L}$, and, hence,~\eqref{eq.Pn} underestimates $\mathrm{P}_\mathrm{N}$. However, we observe that: (i) for $\lambda_{\mathrm{BS}} \in [ 2 \cdot 10^{-4}, 10^{-2} ]$, the overestimation error is smaller than $0.03$), and (ii) for dense scenarios ($\lambda_{\mathrm{BS}} > 10^{-2}$), it never exceeds $0.01$. Generally, we observe that the proposed theoretical model follows the trend of the simulated values.
From Fig.~\ref{fig.f_0}, we also conclude that $\mathrm{P}_\mathrm{L}$ may have a non-trivial minimum. In our scenarios, this is particularly evident when $N_o = 2$.

\begin{figure}[tb]
\centering
\subfloat[$\lambda_{\mathrm{BS}} = 10^{-2}$, $x$-ISD = $\SI{100}{\meter}$]{\label{fig.f_2.1}
	\includegraphics[width=1\columnwidth]{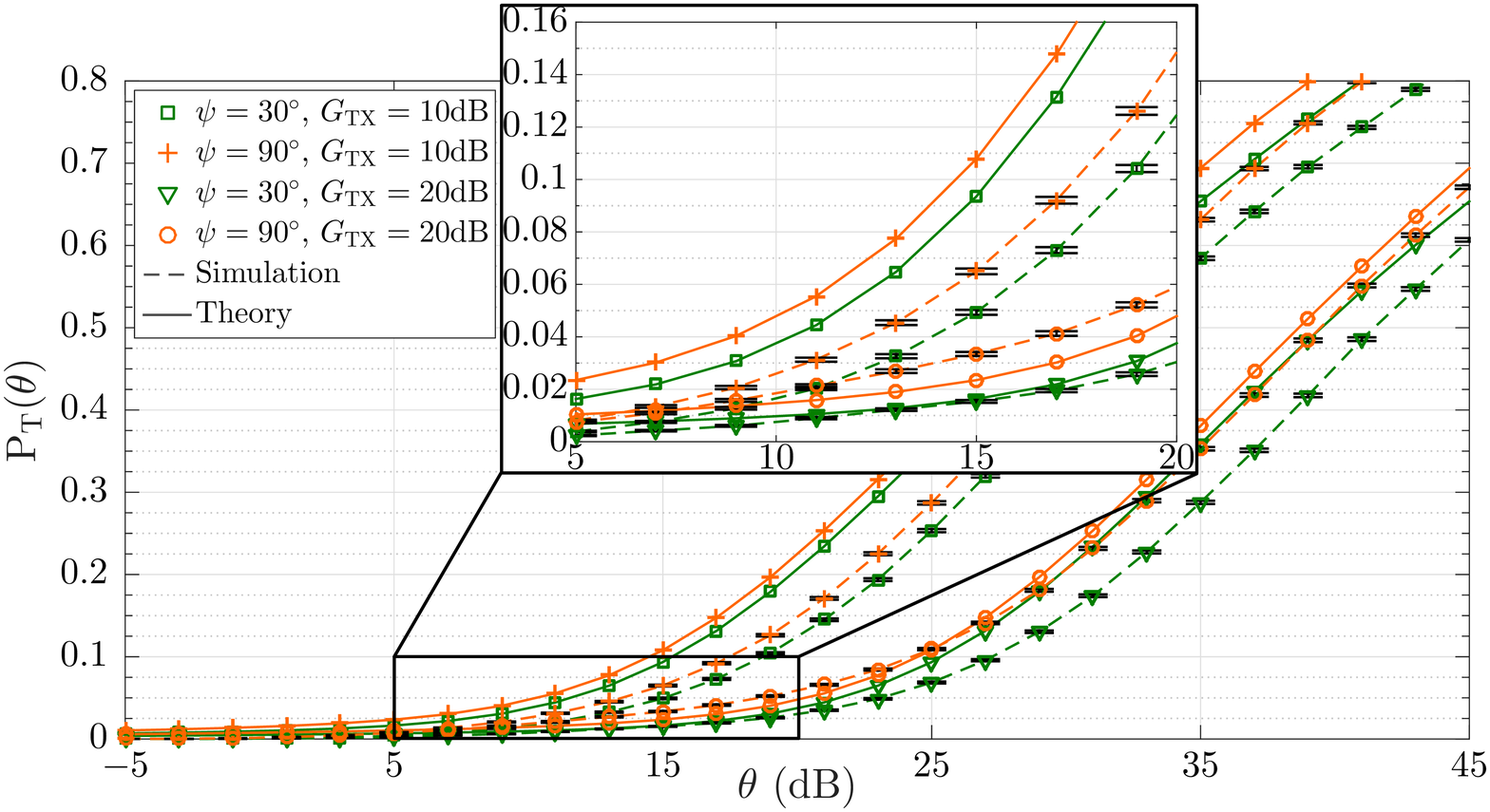}
}\\
\subfloat[$\lambda_{\mathrm{BS}} = 4 \cdot 10^{-3}$, $x$-ISD = $\SI{250}{\meter}$]{\label{fig.f_2.2}
	\includegraphics[width=1\columnwidth]{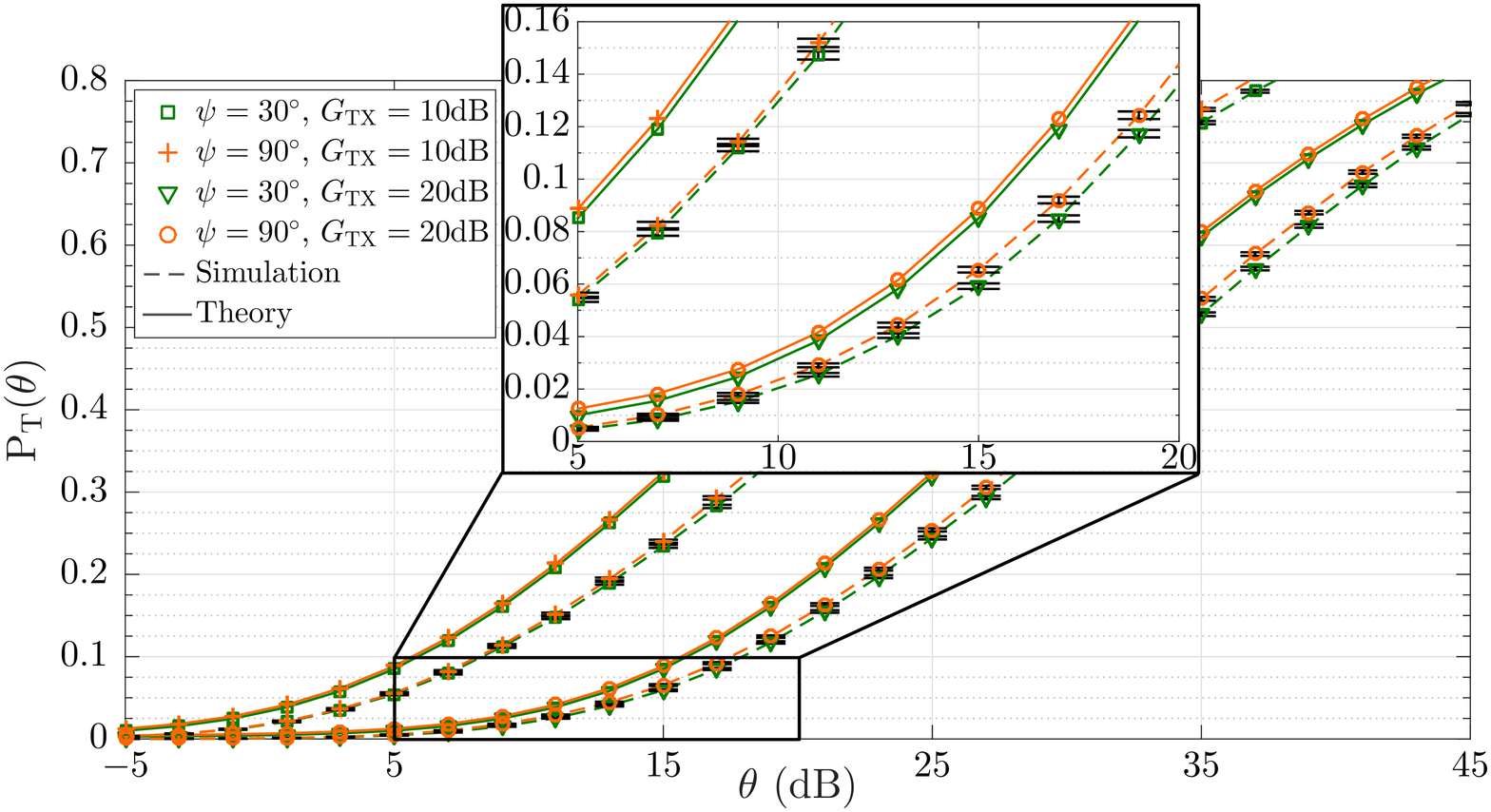}
}
\caption{SINR outage probability $\mathrm{P}_{\mathrm{T}}$ as a function of the threshold $\theta$, for $N_o = 2$, $\alpha_\mathrm{N} = 4$, $\psi = \{30^\circ, 90^\circ\}$ and $G_{\mathrm{TX}} = \{\SI{10}{\dB}, \SI{20}{\dB}\}$.}
\label{fig.f_2}
\end{figure}

\begin{remark}
As we move from sparse to dense scenarios, it becomes more likely for a NLOS BS to be closer to the standard user; thus $\mathrm{P}_\mathrm{L}$ decreases. However, this reasoning holds up to a certain value of density. In fact, at some point, the BS density becomes so high that it becomes increasingly unlikely not to have a LOS BS that is close enough to serve the standard user. This phenomenon may determine a non-trivial minimum in $\mathrm{P}_\mathrm{L}$.
\end{remark}

{The table superimposed to Fig.~\ref{fig.f_0} lists the (simulated) values of $p_\mathrm{L}$, $p_\mathrm{N}$ and the average duration of a blockage event impairing transmissions from BSs on the upper and bottom side of the road. In particular, we observe that a blockage event can occur with a probability greater than $0.22$ and can last up to $\SI{2.68}{\second}$\footnote{{The standard user drives in the East-to-West direction. Hence, the East-to-West blockages have an (average) relative speed equal to \SI{16}{\kilo\meter\per\hour} (namely, $\SI{112}{\kilo\meter\per\hour} - \SI{96}{\kilo\meter\per\hour}$). For blockages with a length equal to $\SI{11.2}{\meter}$ a blockage event is expected to last about $\SI{2.5}{\second}$, which is close to the result of our simulations. The same reasoning applies to West-to-East blockages.}}.}

Fig.~\ref{fig.f_1} shows the effect of the SINR threshold $\theta$ on the outage probability $\mathrm{P}_{\mathrm{T}}(\theta)$, for $N_o = 1$, several antenna beamwidth $\psi$ and a range of BS transmit antenna gains $G_{\mathrm{TX}}$. Here, the vehicular receive antenna gain is set to $G_{\mathrm{RX}} = \SI{10}{\dB}$. In Fig.~\ref{fig.f_1.1}, the $x$-ISD is fixed at \SI{100}{\meter}. It should be noted that the proposed theoretical model, as in Theorem~\ref{th.SINRout}, not only follows the trend of the simulated values of $\mathrm{P}_{\mathrm{T}}(\theta)$ but also it is a tight upper-bound for our simulations for the majority of the values of $\theta$. In addition, the deviation between theory and simulation is negligible when $\theta \in [\SI{-5}{\dB}, \SI{15}{\dB}]$ or $\theta \in [\SI{-5}{\dB}, \SI{10}{\dB}]$, for $G_{\mathrm{TX}} = \SI{10}{\dB}$ or $G_{\mathrm{TX}} = \SI{20}{\dB}$, respectively. On the other hand, that deviation gradually increases as $\theta$ becomes larger. Nevertheless, the maximum Mean Squared Error (MSE) between simulation and theory is smaller than $3.2 \cdot 10^{-3}$. Overall, we observe the following facts:
\begin{itemize}
\item Changing the beamwidth from $30^\circ$ to $90^\circ$ alters the SINR outage probability only by a maximum of $4 \cdot 10^{-2}$. This can be intuitively explained by noting that the serving BS is likely to be close to the vertical symmetry axis of our system model. From Assumption~\ref{ass.antennaPattern}, the standard user aligns its beam towards the serving BS. As such, the values of $J$ and $K$ (see Theorem~\ref{th.LI}) do not largely change on passing from $\psi = 30^\circ$ to $\psi = 90^\circ$. Thus, for the interference component to become substantial, the value of $\psi$ should be quite large.
\item Overall, we observe that when the beamwidth increases, so does $\mathrm{P}_{\mathrm{T}}$. Intuitively, that is because the standard user is likely to receive a large interference contribution via the main antenna lobe.
\item Increasing the value of the maximum transmit antenna gain (from \SI{10}{\dB} to \SI{20}{\dB}) results in a reduction of the SINR outage probability that, for large values of $\theta$, can be greater than $0.25$. This clearly suggests that the increment on the interfering power is always smaller than or equal to the correspondent increment on the signal power. This is mainly because of the directivity of the considered antenna model and the disposition of the BSs.
\end{itemize}

Fig.~\ref{fig.f_1.2} refers to the same scenarios as in Fig.~\ref{fig.f_1.1} except for the $x$-ISD that is equal to \SI{250}{\metre}. In general, we observe that the comments to Fig.~\ref{fig.f_1.1} still hold.
Furthermore, the impact of the value of $\psi$ on $\mathrm{P}_{\mathrm{T}}$ becomes negligible. Intuitively, this can be explained by noting that the number of interfering BSs that are going to be received by the standard user at the maximum antenna gain decreases as $\lambda_{\mathrm{BS}}$ decreases. However, as the BS density decreases (the BSs are more sparsely deployed), it becomes more likely (up to a certain extent) that the number of interfering BSs remains the same, even for a beamwidth equal to $90^\circ$.

Fig.~\ref{fig.f_2} refers to the same scenarios as Fig.~\ref{fig.f_1} with two obstacle lanes on each side of the road ($N_o = 2$). In addition to the discussion for Fig.~\ref{fig.f_1}, we note the following:
\begin{itemize}
\item For the smallest value of the antenna transmit gain ($G_{\mathrm{TX}} = \SI{10}{\dB}$), both the simulated and the proposed  theoretical model produce values of $\mathrm{P}_{\mathrm{T}}$ that are negligibly greater that those when $N_o = 1$.
\item For $x$-ISD = \SI{100}{\meter} and $G_{\mathrm{TX}} = \SI{20}{\dB}$, the SINR outage is slightly greater that the correspondent case as in Fig.~\ref{fig.f_1.1}. In particular, for $\theta \geq \SI{25}{\dB}$, we observe an increment in the simulated $\mathrm{P}_{\mathrm{T}}$ bigger than $9 \cdot 10^{-2}$.
\item As soon as we refer to a sparser network scenario, \mbox{$x$-ISD = \SI{250}{\meter}}, the conclusions drawn for Fig.~\ref{fig.f_1.2} also apply for Fig.~\ref{fig.f_2.2}. Hence, the impact of $\psi$ on $\mathrm{P}_{\mathrm{T}}$ vanishes. 
\end{itemize}

\begin{figure}[t]
\vspace{-3mm}
\centering
\includegraphics[width=1\columnwidth]{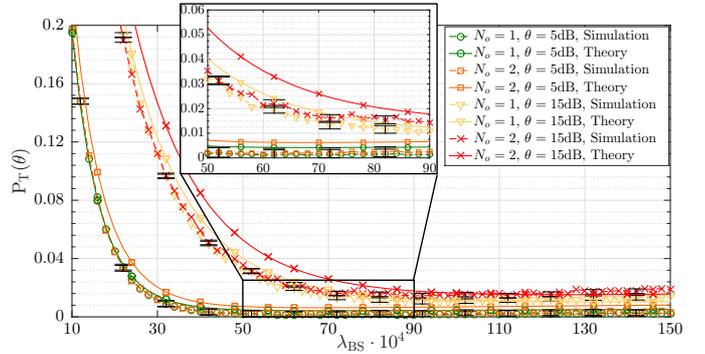}
\caption{SINR outage probability $\mathrm{P}_{\mathrm{T}}$ as a function of the BS density $\lambda_\mathrm{BS}$, for $\theta = \{\SI{5}{\dB},\SI{15}{\dB}\}$ dB, $N_o = \{1, 2\}$, $\alpha_\mathrm{N} = 4$, $\psi = 30^\circ$ and \mbox{$G_{\mathrm{TX}} = \SI{20}{\dB}$}.}
\label{fig.f_4}
\vspace{-3mm}
\end{figure}

\begin{figure}[t]
	\includegraphics[width=1\columnwidth]{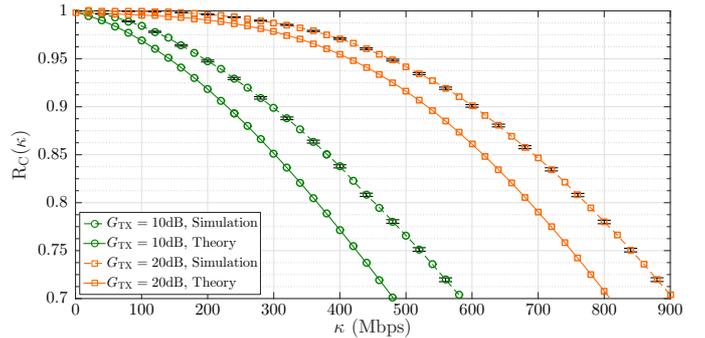}
\caption{Rate coverage probability $\mathrm{R}_\mathrm{C}$ as a function of the threshold $\kappa$, for $\alpha_\mathrm{N} = 4$, $\psi = 30^\circ$, $G_{\mathrm{TX}} = \{\SI{10}{\dB}, \SI{20}{\dB}\}$, $\lambda_\mathrm{BS} = 4 \cdot 10^{-3}$, $N_o = 2$.}
\label{fig.f_3}\vspace{-3mm}
\end{figure}

\begin{figure}[t]
\vspace{-8mm}
\centering
\includegraphics[width=1\columnwidth]{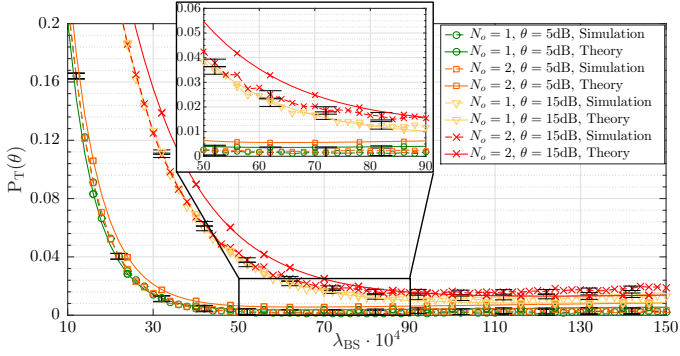}
\caption{SINR outage probability $\mathrm{P}_{\mathrm{T}}$ as a function of the BS density $\lambda_\mathrm{BS}$, for $\theta = \{\SI{5}{\dB},\SI{15}{\dB}\}$ dB, $N_o = \{1, 2\}$, $\alpha_\mathrm{N} = 5.76$, $\psi = 30^\circ$ and \mbox{$G_{\mathrm{TX}} = \SI{20}{\dB}$}. Simulation results obtained for $2R = \SI{20}{\kilo\meter}$.}
\label{fig.f_4_SIM}
\vspace{-5mm}
\end{figure}

From Fig.~\ref{fig.f_1} and Fig.~\ref{fig.f_2}, we already observed that the proposed theoretical model, as in Theorem~\ref{th.SINRout}, follows well the trend of the corresponding simulated values, and it is characterized by an error that is negligible for the most important values of $\theta$ (e.g., $\theta \leq \SI{20}{\dB}$). These facts are further confirmed by Fig.~\ref{fig.f_4}, which shows the value of $\mathrm{P}_{\mathrm{T}}$ as a function of $\lambda_\mathrm{BS}$, for $\theta = \SI{5}{\dB}$ or $\SI{15}{\dB}$, and $\psi = 30^\circ$. In particular, as also shown in Fig.~\ref{fig.f_1} and Fig.~\ref{fig.f_2}, as $\theta$ increases the deviation between the simulations and the theoretical model increases. However, the MSE between theory and simulation never exceeds $5 \cdot 10^{-3}$ in Figs.~\ref{fig.f_2.1} and~\ref{fig.f_2.2}.
{Furthermore, Fig.~\ref{fig.f_4} allows us to expand what was already observed for Fig.~\ref{fig.f_1} and Fig.~\ref{fig.f_2}:
\begin{itemize}
\item As expected, $\mathrm{P}_{\mathrm{T}}$ increases as $N_o$ increases. However, when $N_o$ passes from $1$ to $2$, $\mathrm{P}_{\mathrm{T}}$ increases no more than $1 \cdot 10^{-2}$. Hence, we conclude that the network is particularly resilient to the number of obstacle lanes.
\item The impact of $\lambda_\mathrm{BS}$ on the value of $\mathrm{P}_{\mathrm{T}}$ more evident for sparse scenarios -- $\lambda_\mathrm{BS} \leq 3 \cdot 10^{-3}$ and $\lambda_\mathrm{BS} \leq 5 \cdot 10^{-3}$, for $\theta = \SI{5}{\dB}$ and $\theta = \SI{15}{\dB}$, respectively. Otherwise, the impact of $\lambda_\mathrm{BS}$ is reasonably small, if compared to what happens in a typical bi-dimensional mmWave cellular network~\cite{6932503}. This can be justified by the same reasoning provided for Fig.~\ref{fig.f_1.1}.
\item As the value of $\lambda_\mathrm{BS}$ increases, the interference component progressively becomes dominant again and hence, $\mathrm{P}_{\mathrm{T}}$ is expected to increase. In Fig.~\ref{fig.f_4}, this can be appreciated for $N_o = 2$ and $\theta = \SI{15}{\dB}$.
\end{itemize}
Let us consider again Fig.~\ref{fig.f_4}. In the considered scenarios, it is possible to achieve a value of $\mathrm{P}_{\mathrm{T}}$ smaller than $0.2$ for values of $\lambda_\mathrm{BS} \cong 2.2 \cdot 10^{-3}$.}

Fig.~\ref{fig.f_3} shows the rate coverage probability as a function of the rate threshold $\kappa$, for $\psi = 30^\circ$, $\lambda_\mathrm{BS} = 4 \cdot 10^{-3}$ and \mbox{$N_o = 2$}. From~\eqref{eq.rate}, we remark that the expression of $\mathrm{R}_\mathrm{C}$ directly follows from $\mathrm{P}_{\mathrm{T}}$. For this reason, we observe that the greater the gain $\mathrm{G}_\mathrm{TX}$, the higher the value of $\mathrm{R}_\mathrm{C}$. Finally, we observe that the MSE between simulations and the proposed theoretical approximation is smaller than $5.8 \cdot 10^{-3}$.

{For completeness, our model was validated by considering $\alpha_\mathrm{N} = 5.76$ and a significantly shorter highway section, namely $2R = \SI{20}{\kilo\meter}$.
{We observe that the considered value of $\alpha_\mathrm{N}$ is among the highest NLOS path loss exponent that has ever been measured in an outdoor performance investigation~\cite{6932503}.}
In particular, Fig.~\ref{fig.f_4_SIM} compares simulation and theoretical results for the same transmission parameters and road layout as in Fig.~\ref{fig.f_4}.
The bigger NLOS path loss exponent determines bigger $\mathrm{P}_{\mathrm{T}}$ values than the correspondent cases reported in Fig.~\ref{fig.f_4} (the absolute difference is bigger than $1.1 \cdot 10^{-2}$). Nevertheless, what observed for Fig.~\ref{fig.f_4} applies also for Fig.~\ref{fig.f_4_SIM}. In particular, we conclude that the proposed theoretical model remains valid for shorter road sections.}

\vspace{-5mm}\section{Conclusions and Future Developments}\label{sec.cl}
This paper has addressed the issue of characterizing the downlink performance of a mmWave network deployed along a highway section. In particular, we proposed a novel theoretical framework for characterizing the SINR outage probability and rate coverage probability of a user surrounded by large vehicles sharing the other highway lanes. Our model treated large vehicles as blockages, and hence, they impact on the developed LOS/NLOS model. One of the prominent features of our system model is that BSs are systematically placed at the side of the road, and large vehicles are assumed to drive along parallel lanes. Hence, unlike a typical stochastic geometry system, we assumed that both BS and blockage positions are governed by multiple independent mono-dimensional PPPs that are not independent of translations and rotations. This modeling choice allowed the proposed theoretical framework to model different road layouts.

We compared the proposed theoretical framework with simulation results, for a number of scenarios. In particular, we observed that the proposed theoretical framework can efficiently describe the network performance, in terms of SINR outage and rate coverage probability. Furthermore, we observed the following fundamental properties:
\begin{itemize}
\item Reducing the antenna beamwidth from $90^\circ$ to $30^\circ$ does not necessarily have a disruptive impact on the SINR outage probability, and hence, on the rate coverage probability.
\vspace{-3mm}{\item In contrast with bi-dimensional mmWave cellular networks, the network performance is not largely impacted by values of BS density ranging from moderately sparse to dense deployments.
\item Overall, for a fixed SINR threshold, a reduced SINR outage probability can be achieved for moderately sparse network deployments.}
\end{itemize}

\vspace{-5mm}\appendices\section{Proof of Theorem~\ref{th.LI}}\label{app.A}
For $\mathbb{E}_1 = \{\mathrm{L},\mathrm{N}\}$, the Laplace transform of $\mathrm{I}_{\mathrm{S,E}}$ can be expressed as:
\setlength{\arraycolsep}{0.0em} 
\vspace{-3mm}\begin{eqnarray}
\hspace{-3mm}\mathcal{L}_{\mathrm{I}_{\mathrm{S,E}},\mathbb{E}_1}(s) &{}={}& \mathbb{E}_{\Phi_{\mathrm{S,E}}} \left( \prod_{j \in \Phi_{\mathrm{S,E}}} \mathbb{E}_h \mathbb{E}_\Delta \left( e^{-s |h_j|^2 \Delta_j \ell(r_j)} \right)\right)\label{eq.app.p1.1}\\
&{}\stackrel{(i)}{=}{}& \exp \Bigg(- \mathbb{E}_\Delta \mathbb{E}_h \int_{w(N_o+1)}^{+\infty} (1-e^{-s h \Delta \mathrm{C_E} r^{-\alpha_\mathrm{E}}})\nonumber\\
&& \cdot \, \frac{2 r q \lambda_\mathrm{E}}{\sqrt{r^2 - w^2(N_o+1)^2}} \, dr \Bigg)\label{eq.app.p1.1_}
\end{eqnarray}
where $\mathbb{E}_{\Phi_{\mathrm{S,E}}}$ represents the expectation with respect to the distance of each BS in $\Phi_{\mathrm{S,E}}$ from $O$. Similarly, operators $\mathbb{E}_\Delta$ and $\mathbb{E}_h$ signify the expectation with respect to the overall antenna gain and the small-scale fading gain associated with the transmissions of each BS, respectively.
For the sake of compactness, from $(i)$ onward we refer to $|h|^2$ simply as $h$.
We observe that equality $(i)$ arises from the definition of a probability generating functional (pgfl) of a PPP~\cite[Definition~4.3]{haenggi2013stochastic} and the mapping theorem applied to $\Phi_{\mathrm{S,E}}$~\cite[Theorem~2.34]{haenggi2013stochastic}. In addition, the pgfl allows us to drop the relation to a specific BS $j$ in the terms expressing a distance of a BS to $O$, its channel and antenna gains. For this reason, in the integrand function, we simply refer to terms $r$, $h$ and $\Delta$.

Let $a$ and $b$ be two real numbers greater than or equal to $w(N_o+1)$ and such that $a \leq b$. With regards to~\eqref{eq.app.p1.1_}, we condition with respect to a specific value of $h$ and $\Delta$, and we approximate the following term\footnote{For clarity, we define $[f(x)]_{x=a}^b = f(b) - f(a)$. }:
\setlength{\arraycolsep}{0.0em} 
\begin{eqnarray}
&&\int_{a}^{b} (1-e^{-s h \Delta \mathrm{C_E} r^{-\alpha_\mathrm{E}}}) \frac{2 r q \lambda_\mathrm{E}}{\sqrt{r^2 - w^2(N_o+1)^2}} \, dr\notag\\
&{}\stackrel{(i)}{=}{}& \int_{\sqrt{a^2-w^2(N_o+1)^2}}^{\sqrt{b^2-w^2(N_o+1)^2}} \left(1-e^{-s h \Delta \mathrm{C_E} (t^2+w^2(N_o+1)^2)^{-\alpha_\mathrm{E}/2}}\right) 2 q \lambda_\mathrm{E} \, dt \nonumber\\
&{}\stackrel{(ii)}{\cong}{}& \int_{a}^{b} \left(1-e^{-s h \Delta \mathrm{C_E} t^{-\alpha_\mathrm{E}}}\right) 2 q \lambda_\mathrm{E} \, dt \label{eq.R.relaxed}\\
&{}\stackrel{(iii)}{=}{}& -2 q \lambda_\mathrm{E} \int_{a^{-\alpha_\mathrm{E}}}^{b^{-\alpha_\mathrm{E}}} (1-e^{-s h \Delta \mathrm{C_E} x}) \alpha_\mathrm{E}^{-1} x^{-\alpha_\mathrm{E}^{-1} - 1} \, dx \nonumber\\
&{}\stackrel{(iv)}{=}{}& \overbrace{2 q \lambda_\mathrm{E} \left[(1-e^{-s h \Delta \mathrm{C_E} x}) x^{-\alpha_\mathrm{E}^{-1}}\right]_{x = a^{-\alpha_\mathrm{E}}}^{b^{-\alpha_\mathrm{E}}}}^{\Theta(h,\Delta)}\notag\\
&& \underbrace{-2 q \lambda_\mathrm{E} \int_{a^{-\alpha_\mathrm{E}}}^{b^{-\alpha_\mathrm{E}}} s h \Delta \mathrm{C_E} x^{-\frac{1}{\alpha_\mathrm{E}}} e^{-s h \Delta \mathrm{C_E} x} \, dx}_{\Lambda(h,\Delta)},\label{eq.intApp}
\end{eqnarray}
where $(i)$ arises from the change of variable \mbox{$t \leftarrow \sqrt{r^2 - w^2(N_o+1)^2}$}, while $(ii)$ assumes that $w(N_o+1)$ is equal to $0$ (see Section~\ref{subsec.val} for the validation of the proposed theoretical framework). Equality $(iii)$ arises by applying the changes of variable $y \leftarrow t^{\alpha_{\mathrm{E}}}$ and then $x \leftarrow y^{-1}$. In addition, in $(iv)$, we resort to an integration by parts.

With regards to~\eqref{eq.intApp}, we keep the conditioning to $\Delta$ and calculate the expectation of $\Theta(h,\Delta)$ and $\Lambda(h,\Delta)$, with respect to $h$. From Assumption~\ref{ass.ch}, it should be noted that we refer to a Rayleigh channel model, and, hence, the following relation holds:
\begin{equation}
\!\mathbb{E}_h \left[\Theta(h,\Delta)\right] = 2 q \lambda_\mathrm{E}\! \left[x^{-\alpha_\mathrm{E}^{-1}} \!\!\left(1-\frac{1}{s \Delta \mathrm{C_E} x + 1}\right) \right]_{x = a^{-\alpha_\mathrm{E}}}^{b^{-\alpha_\mathrm{E}}}\!\!\!\!\!\!\!\!\!\!\!\!.\!\!
\end{equation}
Term $\mathbb{E}_h \left[\Lambda(h,\Delta)\right]$ can be found as follows:
\setlength{\arraycolsep}{0.0em} 
\begin{eqnarray}
\mathbb{E}_h \left[\Lambda(h,\Delta)\right] &{}={}& -2 q \lambda_\mathrm{E} \int_{a^{-\alpha_\mathrm{E}}}^{b^{-\alpha_\mathrm{E}}} x^{-\frac{1}{\alpha_\mathrm{E}}} \notag\\
&&  \cdot \int_0^\infty s h \Delta \mathrm{C_E}  e^{-(s \Delta \mathrm{C_E} x+1)h} \, dh \, dx \label{eq.Gauss}\\
&{}={}& -2 q \lambda_\mathrm{E} \int_{a^{-\alpha_\mathrm{E}}}^{b^{-\alpha_\mathrm{E}}}  \!\!\!x^{-\frac{1}{\alpha_\mathrm{E}}} \frac{\partial}{\partial x}\left(-\frac{1}{s \Delta \mathrm{C_E} x + 1}\right) \, dx\nonumber\\
&{}\hspace{-33mm}\stackrel{(i)}{=}{}&\hspace{-16mm}-2 q \lambda_\mathrm{E} (s\Delta \mathrm{C_E})^{\frac{1}{\alpha_\mathrm{E}}} \!\!\int_{-(s \Delta \mathrm{C_E} a^{-\alpha_\mathrm{E}}+1)^{-1}}^{-(s \Delta \mathrm{C_E} b^{-\alpha_\mathrm{E}}+1)^{-1}}  \!\!\!\left(-\frac{1}{t}-1\right)^{-\frac{1}{\alpha_\mathrm{E}}}  dt\nonumber
\end{eqnarray}
\begin{eqnarray}
&{}\stackrel{(ii)}{=}{}&-2 q \lambda_\mathrm{E} (s\Delta \mathrm{C_E})^{\frac{1}{\alpha_\mathrm{E}}} \Bigg[t (-t^{-1})^{-\frac{1}{\alpha_\mathrm{E}}}\Gamma\left(\frac{1}{\alpha_\mathrm{E}} + 1\right)\nonumber\\
&& \cdot {}_2\tilde{F}_1\left(\frac{1}{\alpha_\mathrm{E}}, \frac{1}{\alpha_\mathrm{E}}+1;\frac{1}{\alpha_\mathrm{E}}+2;-t\right)\Bigg]_{t = -(s \Delta \mathrm{C_E} a^{-\alpha_\mathrm{E}}+1)^{-1}}^{-(s \Delta \mathrm{C_E} b^{-\alpha_\mathrm{E}}+1)^{-1}}\nonumber,
\end{eqnarray}
where $(i)$ arises from the change of variable $t \leftarrow -\frac{1}{s \Delta \mathrm{C_E} x + 1}$. Let us signify with ${}_2\tilde{F}_1(a, b; c; z) = \sum_{k = 0}^\infty \frac{\{a\}_k\{b\}_k}{\{c\}_k} \frac{z^k}{k!}$ the Gauss hypergeometric function\footnote{We observe that $z$ is always a real number, which allows us to significantly reduce the complexity of the whole numerical integration process~\cite{hankin:2015}.}. We observe that the integral as in equality $(i)$ is closely related to that as in~\cite[Eq.~(3.228.3)]{jeffrey2007table}, and after some manipulations we have equality $(ii)$.
From the approximation in~\eqref{eq.R.relaxed}, it follows that \mbox{$a \cong \sqrt{a^2-w^2(N_o+1)^2}$} and $b \cong \sqrt{b^2-w^2(N_o+1)^2}$. Hence, we observe that $\mathcal{L}_{\mathrm{I}_{\mathrm{S,E}},\mathbb{E}_1}(s)$, conditioned on the gain $\Delta$ (see~\eqref{eq.sinr_O}), can be expressed as follows:
\setlength{\arraycolsep}{0.0em} 
\begin{eqnarray}
\mathcal{L}_{\mathrm{I}_{\mathrm{S,E}},\mathbb{E}_1}(s) &{}\cong{}& \mathcal{L}_{\mathrm{I}_{\mathrm{S,E}},\mathbb{E}_1}(s;a,b,\Delta)\Big|_{a = 0, b = +\infty}\label{eq.LI}\label{eq.app.pTh1}\\
&&\hspace{-12mm}= \exp \Bigg( - \Big(\mathbb{E}_h [\Theta(h,\Delta)] + \mathbb{E}_h [\Lambda(h,\Delta)]\Big|_{a = 0, b = +\infty}\Big)\Bigg).\nonumber
\end{eqnarray}

Let us focus on the transmit antenna gain of the \mbox{$j$-th} interfering BS, which has a PDF that depends on the distance $r_j$ and the orientation of the beam $\epsilon_i$. To take into account the exact formulation of the BS transmit antenna gain would make the performance model intractable. As such, we instead make the approximation that the transmit antenna gain is always equal to $g_{\mathrm{TX}}$.

\begin{figure}[t]
\vspace{-5mm}
\centering
\includegraphics[width=0.7\columnwidth]{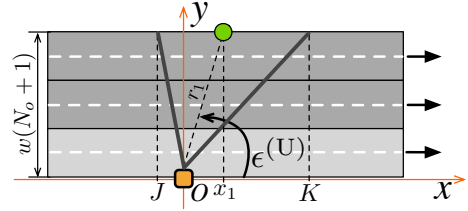}\vspace{-3mm}
\caption{Case where the standard user is served by a BS from the upper side of the road.}\vspace{-2mm}
\label{fig.rxGainExpl}
\vspace{-4mm}
\end{figure}

With regards to~\eqref{eq.app.p1.1}, we observe that after conditioning on the standard user being connected to a BS at a distance $r_1$ from $O$, then the receive antenna gain $g$ of the interfering BSs $j$ is determined by the parameter list $<|x_1|,p,r,\mathrm{\mathbb{S}_1,\mathbb{E}_1,S,E}>$, where: (i) $|x_1|$ is the absolute value of the $x$-axis coordinate of BS $1$, (ii) $p$ captures the fact that the interfering BS is at a location on the positive (right-hand side of $y$-axis, $\mathrm{RX}$) or negative side (left-hand side of $y$-axis, $\mathrm{LX}$) of the $x$-axis, and (iii) $r$ is the distance of the interfering BS to $O$.

Let us consider the $x$-axis coordinates $J$ and $K$ of the points where the two rays defining the antenna beam of the standard user intersect the side of the road, as shown in Fig.~\ref{fig.rxGainExpl}. The receive antenna gain of the interfering BS also depends on: (i) the fact the standard user connects to a BS on the upper/bottom side of the road ($\mathbb{S}_1 \in \{\mathrm{U},\mathrm{B}\}$), that can be in LOS/NLOS ($\mathbb{E}_1 \in \{\mathrm{L},\mathrm{N}\}$) with respect to the standard user, (ii) the values of $\mathrm{S}$ and $\mathrm{E}$, and (iii) the specific configuration of the values of $|x_1|$, $J$ and $K$. By invoking the same approximation as in~\eqref{eq.R.relaxed}, we say that $r \cong \sqrt{r^2 - w^2(N_o+1)^2}$ and the following parameters determine the receiver gain:
\begin{itemize}
\item $\mathbb{S}_1 = \mathrm{U}$, $\mathbb{E}_1 = \mathrm{L}$, $\mathrm{S} = \mathrm{U}$ and $\mathrm{E} = \mathrm{L}$ - we divide this case into the following subcases:
	\begin{itemize}
	\item If the value of $|x_1|$ is such that $J > 0$ - we observe that there are no LOS BSs at a distance smaller than $|x_1|$. Hence, it follows that
	\begin{equation}\label{eq.p1.2}
		g = G_\mathrm{RX} \quad \text{if} \quad \left\{ 
			\begin{array}{l l}
				|x_1| \leq r \leq K & \\
				p = \mathrm{RX} & 
				\end{array} \right.
	\end{equation}
	\begin{equation}
		\hspace{-10mm}g = g_\mathrm{RX} \quad \text{if} \quad \left\{ 
			\begin{array}{l l}
				K \leq r \leq +\infty & \\
				p = \mathrm{RX} & 
				\end{array} \right. \,\text{or}\, \left\{ 
			\begin{array}{l l}
				|x_1| \leq r \leq +\infty & \\
				p = \mathrm{LX} & 
				\end{array} \right.\hspace{-5mm}
	\end{equation}

		\item If $J \leq 0$ - by following the same reasoning as before, in addition to the case as in~\eqref{eq.p1.2}, it follows that
	\begin{equation}\label{eq.s_}
		\hspace{-10mm}g = g_\mathrm{RX} \quad \text{if} \quad \left\{ 
			\begin{array}{l l}
				K \leq r \leq +\infty & \\
				p = \mathrm{RX} & 
				\end{array} \right. \,\text{or}\, \left\{ 
			\begin{array}{l l}
				|J| \leq r \leq +\infty & \\
				p = \mathrm{LX} & 
				\end{array} \right.\hspace{-5mm}
	\end{equation}
	\begin{equation}\label{eq.p1.n}
		g = G_\mathrm{RX} \quad \text{if} \quad \left\{ 
			\begin{array}{l l}
				|x_1| \leq r \leq |J| & \\
				p = \mathrm{LX} & 
				\end{array} \right.
	\end{equation}
	\end{itemize}
	\item $\mathbb{S}_1 = \mathrm{U}$, $\mathbb{E}_1 = \mathrm{L}$, $\mathrm{S} = \mathrm{U}$ and $\mathrm{E} = \mathrm{N}$ - we apply the same reasoning as before by bearing in mind that it is impossible for a NLOS BS to be at a distance that is smaller than $\mathrm{A}_\mathrm{N}(r_1)$ to $O$. Equivalently, it is impossible for a NLOS BS to be associated with a $x$-axis coordinate smaller than $x_\mathrm{N}(r_1) = \sqrt{(\mathrm{A}_\mathrm{N}(r_1))^2 - w^2(N_o+1)^2}$. In particular, for $J \leq 0$, the value of $g$ can be derived as in~\eqref{eq.p1.2} and~\eqref{eq.s_}-\eqref{eq.p1.n}, where term $|x_1|$ is replaced by $x_\mathrm{N}(r_1)$. On the other hand, for $J > 0$, the value of $g$ can be expressed as follows:
    \setlength{\arraycolsep}{0.0em}
	\begin{eqnarray}
		\hspace{-2mm}g = g_\mathrm{RX} \quad &\text{if}& \quad \left\{ 
			\begin{array}{l l}
				x_\mathrm{N}(r_1) \leq r \leq J  \,\,\text{or} \,\,K \leq r \leq +\infty& \\
				p = \mathrm{RX} & 
				\end{array} \right.\hspace{-5mm}\notag\\
			&\text{or}&\quad\left\{ 
			\begin{array}{l l}
				x_\mathrm{N}(r_1) \leq r \leq +\infty & \\
				p = \mathrm{LX} & 
				\end{array} \right.
	\end{eqnarray} 
\begin{equation}
		g = G_\mathrm{RX} \quad \text{if} \quad \left\{ 
			\begin{array}{l l}
				J \leq r \leq K & \\
				p = \mathrm{RX} & 
				\end{array} \right.
	\end{equation}
\item $\mathbb{S}_1 = \mathrm{U}$, $\mathbb{E}_1 = \mathrm{L}$, $\mathrm{S} = \mathrm{B}$ and $\mathrm{E} = \mathrm{L}$ - from Assumption~\ref{ass.antennaPattern}, we observe that $g$ is always equal to $g_\mathrm{RX}$. In addition, we note that it is not possible to have a LOS BS at a distance smaller than $r_1$. Hence, we have only two possible configurations:
\begin{equation}\label{eq.p1.ext1}
		\hspace{-5mm}g = g_\mathrm{RX} \quad \text{if} \quad \left\{ 
			\begin{array}{l l}
				|x_1| \leq r \leq +\infty & \\
				p = \mathrm{RX} & 
				\end{array} \right. \,\text{or}\, \left\{ 
			\begin{array}{l l}
				|x_1| \leq r \leq +\infty & \\
				p = \mathrm{LX} & 
				\end{array} \right.\hspace{-5mm}
\end{equation}

\item $\mathbb{S}_1 = \mathrm{U}$, $\mathbb{E}_1 = \mathrm{L}$, $\mathrm{S} = \mathrm{B}$ and $\mathrm{E} = \mathrm{N}$ - similarly to the previous case, we observe that $g$ is equal to $g_\mathrm{RX}$ and it is not possible to have a NLOS BS at a distance smaller than $x_\mathrm{N}(r_1)$. Hence, we have the following cases:
\begin{equation}\label{eq.p1.ext3}
		\hspace{-5mm}g = g_\mathrm{RX} \,\,\,\, \text{if} \,\,\,\, \left\{ 
			\begin{array}{l l}
				x_\mathrm{N}(r_1) \leq r \leq +\infty & \\
				p = \mathrm{RX} & 
				\end{array} \right. \,\text{or}\, \left\{ 
			\begin{array}{l l}
				x_\mathrm{N}(r_1) \leq r \leq +\infty & \\
				p = \mathrm{LX} & 
				\end{array} \right.\hspace{-1mm}
\end{equation}
\item With regards the remaining parameter combinations where $\mathbb{S}_1 = \mathrm{U}$, $\mathbb{E}_1 = \mathrm{N}$, we observe the following cases:
\begin{itemize}
\item $\mathrm{S} = \mathrm{U}$, $\mathrm{E} = \mathrm{L}$ - we define $x_\mathrm{L}(r_1) = \sqrt{(\mathrm{A}_\mathrm{L}(r_1))^2 - w^2(N_o+1)^2}$. If $x_\mathrm{L}(r_1) > K$, refer to~\eqref{eq.p1.ext1} and replace $|x_1|$ with $x_\mathrm{L}(r_1)$. Otherwise, refer to~\eqref{eq.p1.2}-\eqref{eq.p1.n} and replace $|x_1|$ with $x_\mathrm{L}(r_1)$.
\item $\mathrm{S} = \mathrm{U}$, $\mathrm{E} = \mathrm{N}$ - refer to~\eqref{eq.p1.2}-\eqref{eq.p1.n}.
\item $\mathrm{S} = \mathrm{B}$, $\mathrm{E} = \mathrm{L}$ - refer to~\eqref{eq.p1.ext1} and replace $|x_1|$ with $x_\mathrm{L}(r_1)$.
\item $\mathrm{S} = \mathrm{B}$, $\mathrm{E} = \mathrm{N}$ - refer to~\eqref{eq.p1.ext1}.
\end{itemize}
\item By following the above approach, it is possible to derive all the remaining configurations. In particular, the characterization of $g$, for a parameter configuration where $\mathbb{S}_1 = \mathrm{B}$ and $\mathrm{S} = \mathrm{B}$ ($\mathbb{S}_1 = \mathrm{B}$ and $\mathrm{S} = \mathrm{U}$) follows exactly the same rule of the corespondent parameter list, where $\mathbb{S}_1 = \mathrm{U}$ and $\mathrm{S} = \mathrm{U}$ ($\mathbb{S}_1 = \mathrm{U}$ and $\mathrm{S} = \mathrm{B}$).
\end{itemize}
The aforementioned parameter configurations are also summarized in Table~\ref{tab.1}. With regards to parameter $p$, we observe that the probability $\mathbb{P}[p]$ of $p$ being equal to $\mathrm{DX}$ or $\mathrm{RX}$ is $0.5$. 

Consider~\eqref{eq.LI}, all the elements are in place to explicitly calculate the expectation with respect to $\Delta$. In particular, it follows that $\mathcal{L}_{\mathrm{I}_{\mathrm{S,E}}}(s)$ can be expressed as: 
\setlength{\arraycolsep}{0.0em}
\begin{eqnarray}
	\mathcal{L}_{\mathrm{I}_{\mathrm{S,E},\mathbb{E}_1}}(s) \hspace{-0.5mm}&{}\stackrel{(i)}{\cong}{}& \hspace{-0.5mm}\exp \left(\hspace{-1mm} - \mathbb{E}_\Delta\hspace{-1mm}\left(\mathbb{E}_h [\Theta(h,\Delta) \hspace{-1mm}+\hspace{-1mm} \Lambda(h,\Delta)]\Big|_{a = 0,b = +\infty}\right)\hspace{-0.1mm}\right)\nonumber
\end{eqnarray}
\begin{eqnarray}	
&{}\hspace{-2mm}\stackrel{(ii)}{\cong}{}{}& \exp \left( - \hspace{-3mm}\mathop{\sum_{\mathbb{S}_1\in\{\mathrm{U,B}\}}}_{(a,b,\Delta) \in \mathcal{C}_{\mathrm{|x_1|,\mathbb{S}_1,\mathbb{E}_1,S,E}}} \hspace{-7.5mm}\mathbb{P}[p]\left(\mathbb{E}_h [\Theta(h,\Delta) + \Lambda(h,\Delta)]\Big|_{a,b,\Delta}\right)\hspace{-1mm}\right)\nonumber\hspace{-2mm}\\
&=& \hspace{-4mm}\mathop{\mathop{\mathop{\prod_{\mathbb{S}_1\in\{\mathrm{U,B}\},}}}_{(a,b,\Delta) \in \mathcal{C}_{\mathrm{|x_1|,\mathbb{S}_1,\mathbb{E}_1,S,E}}}} \hspace{-10mm} \hspace{-1mm}\exp \left( \hspace{-1mm}- \frac{1}{2}\left(\hspace{-1mm}\mathbb{E}_h [\Theta(h,\Delta)]\Big|_{a,b,\Delta} \hspace{-3mm}+ \mathbb{E}_h [\Lambda(h,\Delta)]\Big|_{a,b,\Delta}\right)\hspace{-1mm}\right)\nonumber\hspace{-2mm}\\
&=&	\mathop{\mathop{\mathop{\prod_{\mathbb{S}_1\in\{\mathrm{U,B}\},}}}_{(a,b,\Delta) \in \mathcal{C}_{\mathrm{|x_1|,\mathbb{S}_1,\mathbb{E}_1,S,E}}}} \sqrt{\mathcal{L}_{\mathrm{I}_{\mathrm{S,E}},\mathbb{E}_1}(s;a,b,\Delta)},
\end{eqnarray}
where $(i)$ is~\eqref{eq.LI}. From the previous discussion, for a given $|x_1|$, $\Delta$ can either be equal to $g_\mathrm{TX}g_\mathrm{RX}$ or $g_\mathrm{TX}G_\mathrm{RX}$. In particular, the value of $\Delta$ is determined by the list of parameters \mbox{$<\mathrm{|x_1|,\mathbb{S}_1,\mathbb{E}_1,S,E}>$}, where terms $a$ and $b$ are the minimum and maximum distance $r$ to an interfering BS, respectively. We define sequence $\mathcal{C}_{\mathrm{|x_1|,\mathbb{S}_1,\mathbb{E}_1,S,E}}$. This sequence consists of all the possible parameter configurations $(a,b,\Delta)$. For instance, if $\mathbb{S}_1 = \mathrm{U}$, $\mathbb{E}_1 = \mathrm{L}$, \mbox{$\mathrm{S} = \mathrm{U}$}, \mbox{$\mathrm{E} = \mathrm{L}$} and \mbox{$J > 0$}, sequence $\mathcal{C}_{\mathrm{|x_1|,\mathbb{S}_1,\mathbb{E}_1,S,E}}$ consists of: $(|x_1|,K,g_\mathrm{TX}G_\mathrm{RX})$, $(K,+\infty,g_\mathrm{TX}g_\mathrm{RX})$ and $(|x_1|,+\infty,g_\mathrm{TX}g_\mathrm{RX})$. We note that each element of a sequence $\mathcal{C}_{\mathrm{|x_1|,\mathbb{S}_1,\mathbb{E}_1,S,E}}$ occurs with probability $\mathbb{P}[p]$. Furthermore, we observe that like those in~\eqref{eq.p1.ext1} and~\eqref{eq.p1.ext3}, sequence $\mathcal{C}_{\mathrm{|x_1|,\mathbb{S}_1,\mathbb{E}_1,S,E}}$ lists twice the same parameter configuration. Given these reasons, the term \mbox{$\mathbb{E}_\Delta(\mathbb{E}_h [\Theta(h,\Delta) + \Lambda(h,\Delta)]\big|_{a = 0,b = +\infty})$} can be approximated as in $(ii)$. After some manipulations of $(ii)$, we get to~\eqref{eq.Th1}, which concludes the proof.

\bibliographystyle{IEEEtran}
\bibliography{IEEEabrv,bib}

\end{document}